\documentclass{aa}
\usepackage{threeparttable}
\usepackage[varg]{txfonts}
\usepackage{natbib}
\usepackage{graphicx}
\usepackage{float}
\usepackage{lscape}
\usepackage{caption}
\usepackage{txfonts}
\usepackage{lipsum}
\usepackage{subcaption} 
\usepackage{lscape}                                 
\usepackage{placeins}                                      
\usepackage{sidecap}
\usepackage{xcolor}
\usepackage[utf8]{inputenc}
\usepackage{threeparttable}
\titlerunning{}
\definecolor{myblue}{rgb}{0.0, 0.0, 1.0}
\usepackage{hyperref}
\hypersetup{
    colorlinks=true,
    linkcolor=myblue,
    citecolor=myblue,
    urlcolor=myblue
}
\usepackage{comment}
\usepackage{multicol}
\titlerunning{}
\begin{document} 
\defcitealias{PapVII}{Paper VII}
\defcitealias{PapVIII}{Paper VIII}
\defcitealias{PapIX}{Paper IX}
\defcitealias{PapV}{Paper V}
\defcitealias{PapI}{Paper I}
\defcitealias{PapIV}{Paper IV}

   \title{The VISCACHA survey}
   \subtitle{XIV. Chemical evolution history of the SMC : The Southern Bridge Clusters.}
   \author{Saroon S
          \inst{1}\fnmsep\thanks{saroonsasi19@gmail.com}
          \and
           Dias B.\inst{1}
           \and
           Tsujimoto T.\inst{2}
           \and
           Maia F.\inst{3}
           \and
           Ferreira B. P. L.\inst{4}
           \and
           Oliveira R. A. P. \inst{5}
           \and 
           Parisi M. C. \inst{6,7}
           \and
           Pérez-Villegas A.\inst{8} 
           \and
           Minniti D.\inst{1,9}
           \and
           De Bortoli B.J. \inst{10,11}
           \and
           Bica E.\inst{12}
           \and
           Westera P.\inst{13}
           \and
           Katime Santrich O.J.\inst{14}
           \and
           Geisler D.\inst{15,16}
           \and
           David Sanmartim\inst{17}
           \and
           Bruno Correa Quint\inst{17}
           \and
           Luciano Fraga\inst{18}
           \and
           J. F. C. Santos Jr.\inst{4}
           \and
           Garro E.R.\inst{19}
           \and
           Jos\'e G. Fern\'andez-Trincado\inst{20}
           \and
           Casmir O.O.\inst{20}
           }
          
   \institute{
   $^{1}$ Instituto de Astrof\'isica, Departamento de F\'isica y Astronom\'ia, Facultad de Ciencias Exactas, Universidad Andres Bello, Fernandez Concha, 700, Las Condes, Santiago, Chile.  \\ 
   $^{2}$ National Astronomical Observatory of Japan, Mitaka, Tokyo 181-8588, Japan.\\
   $^{3}$ Instituto de F\'isica - Universidade Federal do Rio de Janeiro, Av. Athos da Silveira Ramos, 149, Rio de Janeiro, 21941-909, Brazil.\\
   $^{4}$Departamento de F\'isica, ICEx - Universidade Federal de Minas Gerais, Av. Antônio Carlos 6627, Belo Horizonte 31270-901, Brazil.\\
   $^{5}$Astronomical Observatory, University of Warsaw, Al. Ujazdowskie 4, 00-478 Warszawa, Poland.\\
   $^{6}$Observatorio Astronómico, Universidad Nacional de Córdoba, Laprida 854, X5000BGR, Córdoba, Argentina.\\
   $^{7}$Instituto de Astronomía Te\'orica y Experimental (CONICET-UNC), Laprida 854, X5000BGR, Córdoba, Argentina.\\
   $^{8}$ Instituto de Astronomıa, Universidad Nacional Aut\'onoma de M\'exico, Apartado Postal 106, C. P. 22800, Ensenada, B. C., Mexico.\\
   $^{9}$ Vatican Observatory, V00120 Vatican City State, Italy\\
   $^{10}$ Facultad de Ciencias Astronómicas y Geofísicas, Universidad Nacional de La Plata, Paseo del Bosque s/n, 1900 La Plata, Argentina.\\
   $^{11}$ Instituto de Astrofísica de La Plata (CCT La Plata, UNLP-CONICET), Paseo del Bosque s/n, 1900 La Plata, Argentina.\\
   $^{12}$ Universidade FEderal Do Rio Grande do Sul,DEpartamento de AStronomia CP15051, Port Alegre, 91501-970, Brazil.\\
   $^{13}$ Universidade Federal do ABC, Centro de Ciências Naturais e Humanas, Avenida dos Estados, 5001, 09210-580, Brazil.\\
   $^{14}$ Universidade Estadual de Santa Cruz (UESC), Departamento de Ciências Exatas, Rodovia Jorge Amado km 16, 45662-900 Ilhéus,Brazil.\\
   $^{15}$ Departamento de Astronomia, Casilla 160-C, Universidad de Concepcion, Chile.\\
   $^{16}$ Departamento de Astronomía, Facultad de Ciencias, Universidad de La Serena. Av.Raul Bitran 1305, La Serena, Chile.\\
   $^{17}$ NSF NOIRLab/NSF–DOE Vera C. Rubin Observatory HQ, 950 N. Cherry Ave., Tucson, AZ 85719, USA (AURA Staff).\\
   $^{18}$ Laborat\'orio Nacional de Astrof\'isica LNA/MCTI, 37504-364 Itajub\'a, MG, Brazil\\
   $^{19}$ ESO - European Southern Observatory, Alonso de Cordova 3107, Vitacura, Santiago, Chile.\\
   $^{20}$ Instituto de Astronomía, Universidad Católica del Norte, Av. Angamos 0610, Antofagasta, Chile.\\
    } 
   \date{Received---- ; accepted----}
  \abstract
{The chemical evolution history of the Small Magellanic Cloud (SMC) is complex and is best understood through a comprehensive analysis of star clusters across its various regions. The VISCACHA survey aims to fully explain the chemical evolution of SMC star clusters, by analyzing different sub-regions adopted from an existing framework. 
The West Halo (WH) region, which contains the oldest and most metal-poor stellar populations, exhibits a clear age-metallicity relation (AMR) with minimal dispersion. This region reveals a significant dip of $\sim0.5$~dex in metallicity, approximately 6 Gyr ago. This is likely caused by a major merger event that subsequently accelerated the star formation rate. Clusters in the Southern Bridge (SB) and Northern Bridge regions of the SMC may have experienced distinct chemical enrichment histories as suggested by our previous works, but with limited data coverage. Furthermore the AMR of Wing/Bridge (W/B) shows no sign of enrichment caused by the aforementioned merger event, but exhibits signatures of the recent collisions between the clouds, contemporaneous with the epochs of the Magellanic Stream and Bridge formations. 

In this study, we present an updated AMR for the SB region, based on a sample that includes approximately 67\% of its known clusters. Contrary to the expectation of a very unique chemical evolution history, these SB clusters show a trend similar to the one of the WH clusters.
The chemical evolution models that best fit the AMR trend of the SB clusters show excellent agreement with the major merger model proposed for the WH clusters. Building on this, we suggest a new unified chemical evolution model for both the WH and SB clusters, which can be explained by a major merger at $\sim6$~Gyr followed by episodic chemical enrichment over time.

 }
    \keywords{galaxies: dwarf -- galaxies: interactions -- galaxies: Magellanic Clouds}
    \maketitle
\section{Introduction}
Large Magellanic Cloud (LMC) and the Small Magellanic Cloud (SMC), are nearby dwarf irregular galaxies that provide unique opportunities to study various aspects of galactic and stellar evolution. These galaxies are rich in gas and young stars, making them excellent sites for observing ongoing star formation and nucleosynthesis processes.
The LMC and the SMC, positioned at distances of $49.59 \pm 0.09$~kpc (\citealp{Pietrzy2019}) and $62.44 \pm 0.47$~kpc (\citealp{Graczyk2020}) from the Sun, are well within the Milky Way (MW) halo and this proximity allows detailed observations with high resolution. Both galaxies exhibit clear signs of interaction, and substructures like the Magellanic Bridge (\citealp{1961+Hindman,stanimirovic+99}), the Magellanic Stream (MS, \citealp{mathewson+74}) and the Leading Arm (LA, \citealp{1998+Putman}) are the most notable manifestations of these past interactions \citep[ see also][]{nidever+10,Diaz&Bekki2012,donghia+16,PapIII}.

These interactions are particularly significant, since mergers among dwarf galaxies are anticipated to be frequent and serve as key drivers of galaxy evolution, at low and high redshift \citep[e.g.][]{stierwalt+15,poggianti+16,Garavito+21}. 
Precise measurements of stellar proper motions indicate that the LMC and SMC are likely on their first infall into the MW halo (see e.g. \citealp{Besla2007,kallivayalil+13}). However, \citet{2024Vasilev} suggested the possibility of a second infall scenario with the first pericentre passage 5 - 10 Gyr ago. 
Large-scale structures such as the MS can be reproduced by numerical simulations with completely different scenarios, such as, bound \citep{Diaz&Bekki2012} and first infall \citep{Besla2007}. Therefore, to better constrain the evolutionary history of the Magellanic Clouds (MCs), it is essential to analyze smaller-scale structures, along with kinematical and chemical information. 
\citet{nidever+20} showed that the $\alpha -$ knee position of the MCs are lower than that of the less massive MW satellites like Fornax, Sculptor, and Sagittarius suggesting that they were formed in a low density environment and later become part of the MW potential, which supports the first infall scenario.
The challenge of resolving individual stellar populations at considerable distances in most dwarf galaxies accentuates the value of studying the MCs as probes for dwarf-dwarf galaxy interactions, because currently available instrumentation can study even the oldest individual stars from the MCs in survey mode( e.g. \citealp{cioni+11,nidever+17,blanton+17,PapI}).

Chemical evolution refers to the changes in a galaxy's chemical composition over time, driven primarily by processes such as star formation, supernova explosions, and gas inflows and outflows.
Star clusters serve as key tracers of their host galaxy's chemical evolution. Formed from the same material, they preserve a record of the conditions at their formation. They present the great advantage of being able to derive excellent ages, needed to date past events.
Star clusters analyzed as a population can reveal the signatures of past dynamical and chemical evolution history that their host galaxies have undergone, as these effects are imprinted in their chemical, structural, kinematical, and spatial properties \citep[e.g.][]{santos+20, nidever+20, PapIV, deBortili2022, rodriguez+23}. 

The VIsible Soar photometry of star Clusters in tApii and Coxi HuguA (VISCACHA) survey team has been meticulously conducting observations of star clusters in the outskirts of the MCs with the aim of comprehensively analyzing their structures, kinematics, dynamics, and chemical evolution.
The VISCACHA survey uses a comprehensive framework introduced by \citet{Dias+16} aimed at unraveling the complexities of the SMC's chemical evolution puzzle. This systematic approach involves categorizing the SMC cluster population into distinct groups, delineated as the main body, wing and bridge, counter-bridge, 
and west halo (WH).
\begin{figure}[t]
    \includegraphics[width=\columnwidth]{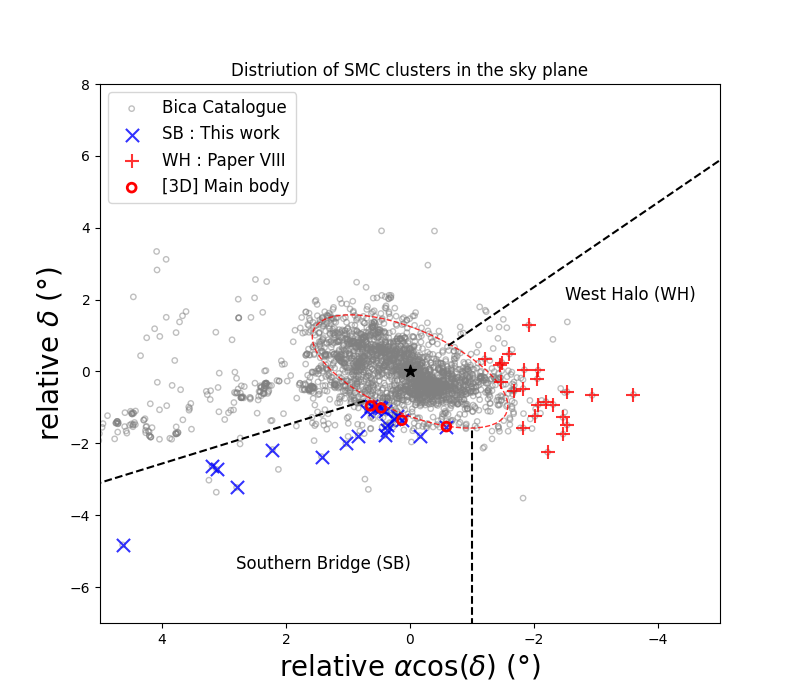}
    \caption{The 2-Dimensional (2D) distribution of SMC clusters analyzed in this work (blue crosses), along with those previously examined in \citetalias[][]{PapVIII} and references therein (red crosses), are shown in comparison with the objects cataloged by \cite{Bica+20} as grey circles. The ellipse represents the SMC [2D] main body (explained in section \ref{survey}) and the dashed lines show the boundaries of these subregions as in \citealp{Dias+16}, and \citetalias{PapVII}. The red open circles represents the SB clusters that are part of the main body based on the 3D classification \citepalias[see][ for more details]{PapIV}.}
    \label{SMC}
\end{figure}
In this context, \citet[][hereafter Paper VII]{PapVII} analyzed the W/B region of the SMC, and show there occurred a metallicity dip about 1.5–2 Gyr ago, which coincides with the LMC-SMC-MW interaction forming the MS \citep[e.g.][]{nidever+10}, and another one around 200\,Myr ago, which is about the formation epoch of the Magellanic Bridge \citep[e.g.][]{zivick+18}. 
\citet[][hereafter Paper VIII]{PapVIII} calculated the most updated Age-Metallicity Relation (AMR) for the SMC WH clusters, unveiling a metallicity dip of approximately 0.5 dex around 6 Gyr ago. They modified the chemical evolution models from \citet{Tsujimoto&Bekki_2009}, explored various scenarios, and identified that a major merger with a metal-poor gas cloud or dwarf galaxy, boasting a mass of around $10^8~M_\odot$ (with a mass ratio of 1:4), provides the most plausible explanation for this observed AMR.  
\citet[][hereafter Paper IX]{PapIX} extended this analysis to the Northern Bridge (NB) and Southern Bridge (SB) clusters, using spectroscopic data from the VISCACHA survey. Despite the low number statistics, the study identified that the NB and SB may have undergone distinct chemical enrichment histories.
All of these aforementioned works point in the direction that each SMC region has a unique chemical evolution history. 
\citet{Narloch2021} analyzed 35 SMC star clusters deriving their AMR (see their Fig.10) and found that the oldest clusters align with the major merger model of \citet{Tsujimoto&Bekki_2009}, while intermediate-age clusters support the burst in the \citet{PT98} model. Clusters younger than 1~Gyr show a wide metallicity spread, consistent with our findings. 
These results confirm that the SMC's chemical evolution cannot be adequately described by a single model, highlighting the effectiveness of a region-by-region approach to disentangle its complex evolutionary history. 

This study aims to derive the most up-to-date and complete AMR of the SMC clusters in the SB region to solve another piece of the SMC chemical evolution puzzle. In addition, we re-analyze the parameters of the WH clusters to ensure a consistent methodology across both regions. This enables the construction of a homogeneous AMR dataset for WH and SB clusters, which is then used to test unified chemical evolution scenarios. This paper is structured as follows. In Section 2, we discuss the data and the statistical decontamination procedure. Section 3 describes the methodology used to derive the cluster parameters. In Section 4, we present the results from isochrone fitting and compare them with previous studies. Section 5 discusses the chemical evolution models of the SB clusters, in the context of the AMR. Finally, we summarize our findings and present our conclusions in Section 6.

\section{Photometry of star clusters}
\subsection{Survey overview} \label{survey}
The VISCACHA survey uses the 4.1 m Southern Astronomical Research (SOAR) telescope. The SOAR Adaptive Optics Module Imager (SAMI, \citealp{tokovini+16}) camera, having a Ground Layer Adaptive Optics (GLAO) module helps in achieving deep photometry with a spatial resolution of up to $\sim0.3"$ in the $I$ band. The VISCACHA survey has consistently generated accurate color-magnitude diagrams (CMDs) for clusters in the outskirts of the LMC and the SMC, and in the Magellanic Bridge. Notably, this includes even the oldest, compact clusters nestled within the densest fields of the MCs, a feat often challenging for larger surveys with lower resolution \citep{PapI, dias+20}. The SAMI's field of view spans $3.1' \times 3.1'$, effectively encompassing the majority of clusters to be well beyond their tidal radii.
For data processing, bias and flat-field corrections were applied using standard methods through IRAF. The entire process of data reduction, point-spread function (PSF) photometry, and calibration adhered to the established procedures outlined in \cite{PapI}. Photometric calibration specifically utilized standard-star fields from \cite{Stetson_2000}. 
The sub-regions of the SMC, as defined by \citet{Dias+16}, are characterized by an ellipse centered on the SMC center ($13.19^\circ , -72.8^\circ3$) with an eccentricity of 0.87 and a position angle of $45^\circ$.

Fig.\ref{SMC} shows the 2D distribution of clusters in the SMC plane. The grey open circles represent all the objects cataloged in \cite{Bica+20}. The red plus signs show the WH clusters, whereas the blue crosses represent the SB clusters. Both samples are analyzed in this work, homogeneously deriving their astrophysical parameters (as explained in Section \ref{siesta}). 
Furthermore, we applied the coordinate transformation equations from \citet{MarelandCioni2001} to calculate the 3D positions and distances of the clusters relative to the SMC center. Clusters located within 4 kpc of the SMC center were classified as belonging to the main body of the SMC, following the approach adopted in previous VISCACHA papers (\citealp{PapIII}; \citetalias{PapIX}). The clusters that are part of the main body based on the aforementioned 3D classification are marked with red open circles in Fig. \ref{SMC}.

For the current analysis, we also utilized data from the \textit{Gaia} and SMASH surveys to compare the isochrone fits across multiband CMDs (eg: \citetalias{PapVIII}). 
The \textit{Gaia} Data Release 3 (GDR3; \citealp{2023GaiaDR3}) catalogs were subjected to quality cuts, selecting sources with astrometric excess noise $\leq 1.3$ mas and a re-normalized unit weight error (RUWE) value $< 1.2$. 
The Survey of the MAgellanic Stellar History (SMASH; \citealp{nidever+17}) employed the Dark Energy Camera to survey 480 square degrees of the sky, achieving depths down to 24th magnitude in the ugriz bands. The comparison of our resultant isochrones with SMASH and \textit{Gaia} CMDs are given in the appendix \ref{APPB}. 

\subsection{Statistical decontamination}
Decontaminating these clusters is a crucial step in the study, given the influence of field star contamination. While one approach involves identifying cluster members through proper motions from \textit{Gaia} (\citealp{2023GaiaDR3}), it's noteworthy that the \textit{Gaia} magnitude limit is approximately G $<21$, which is just down-to the Main Sequence Turn-Off (MSTO) magnitude of a $\sim$3~Gyr old SMC cluster. 
An alternative strategy for distinguishing cluster members from field stars involves utilizing photometric data through statistical comparisons of star samples extracted from the cluster region and an offset field. 
The aforementioned statistical decontamination procedure was executed using a method adapted from \cite{Maia2010}. 

A brief explanation of the decontamination procedure is as follows. A cluster field and a nearby comparison field, characterized by comparable field density and reddening, were carefully chosen for each sample. The cluster's radius was determined based on the catalog provided by \cite{Bica+20}.
The CMDs for both the selected cluster and field stars are partitioned into cells of sufficiently small size to discern local variations in field-star contamination across various sequences in the CMD, while still containing a substantial number of stars. The average cell sizes for each cluster are determined by assessing the dispersion of the data across each magnitude/color, resulting in typical values of $\Delta V = 0.75$ mag and $\Delta (V-I) = 0.33$ mag. Membership probability is then assigned based on the overdensities of the cluster stars in relation to the field stars. This is determined using the relation $P = (N_{clu} - N_{fld})/N_{clu}$, where $N_{clu}$ and $N_{fld}$ are the number densities of cluster stars and field stars, respectively in an identical cell on the CMD. Note that when $N_{fld}>N_{clu}$ a null probability assigned instead of negative value. Furthermore, all stars outside the assumed cluster radius are also given a null probability. The same process is done varying position and size of the cells by $1/3$ of its original size, in all possible combinations. The final selection of the most probable cluster population is done based on the average, median and mode of the membership probability obtained in each possible configuration.
This methodology has been extensively employed successfully in previous VISCACHA works (eg: \citetalias{PapVII,PapVIII}) and applied in the analysis of Galactic (\citealp{Angelo+18}) and MCs clusters (\citealp{Maia+2014}). For more details about the membership probability method, refer to \cite{Maia2010}.
\section{Isochrone Fitting Technique} \label{siesta}
In the decontaminated $V$, $V-I$ CMDs of the sample clusters, distinct features, including well-defined Main Sequences, MSTOs, and Red Giant Branches (RGBs), and Red Clumps (RCs) are prominently visible. These CMD characteristics offer crucial insights into the cluster's global physical properties such as age, metallicity, distance, and reddening. These properties can be retrieved by fitting isochrones to the CMDs as shown in our previous works (see; \citetalias{PapVII,PapVIII,PapIX}).

PARSEC isochrones \citep{Bressen2012} are fitted to the decontaminated CMDs, deriving more accurate determinations of age, metallicity, distance, and reddening. 
We used an updated version of the publicly available \texttt{SIESTA} (Statistical matchIng between rEal and Synthetic STellar populAtions; \citealp{Bernardo+2024}) code to obtain the ages, metallicities, distances, color excesses, and binary fractions of the sample clusters. The code functions by comparing the Hess diagram of a given cluster with the corresponding diagrams from synthetic populations, produced from a set of PARSEC+COLOBRI isochrones \citep[version 3.7;][]{Bressen2012,marigo2017new}. Synthetic populations are produced in a four-stage process. First, the isochrone is sampled using random masses drawn from the \citet{kroupa2001} initial mass function. The next step is to emulate non-resolved binaries by selecting a fraction of the population and adding the flux of a secondary star, uniformly sampled from the isochrone, with a minimum mass ratio of 60\%. Following that, synthetic stars are randomly removed from the population, using rejection sampling so that the final luminosity function then matches the one observed for the cluster. Finally, photometric noise is added to the sampled magnitudes, using the error curves for each observed band (here, V and I).

 Hess diagrams are constructed as 2D histograms, with bin widths determined manually to constrain critical features of the CMDs. For this sample of clusters, this corresponds to the MSTOs and the RCs. The comparison between synthetic and observed data is performed in a Bayesian approach using the likelihood function from \citet{tremmel2013modeling} (Eq. \ref{tremmel}) and \textit{priors}. In the updated version of the code, the likelihood is weighted to add more relevance to the brighter stars since the lower main sequence can dominate the likelihood simply by possessing more bins in the Hess diagram than other regions such as the RCs, the RGB and the turn-off. Weights are defined so that all bins brighter than a certain magnitude appear in the likelihood expression as having the same number of terms as those dimmer than the same magnitude. \\
The Tremmel likelihood is given as :
\begin{equation}
\label{tremmel}
\begin{split}
    \ell(\text{CMD}|\text{params}) &= 
    \prod_{\text{col,mag}}\ell_{\text{col,mag}} \\
    &= \prod_{\text{col,mag}}  \frac{\Gamma(0.5 + n_{\text{col,mag}}^{\text{obs}}+n_{\text{col,mag}}^{\text{synt}})    }{
    \Gamma(1+n_{\text{col,mag}}^{\text{obs}})\Gamma(0.5+n_{\text{col,mag}}^{\text{synt}})
    }
\end{split}
\end{equation}

The weighted likelihood is defined as:
\begin{equation}
    \mathcal{L}(\text{CMD}|\text{params}) = \prod_{\text{col,mag}}\ell_{\text{col,mag}}^{w(\text{mag})} 
    \label{eq:weighed_likelihood}
\end{equation}
 The weight function $w(\text{mag})$ is defined using the magnitude of the bins as follows:
 \begin{equation}
 \small
    w(\text{mag}) \equiv w(V) = \left\{
    \begin{array}{lcl}
        1 &,& V>V_{\text{ref}}, \\
        \frac{ \text{Number of non-empty bins with $V>V_{\text{ref}}$ } }{\text{Number of non-empty bins with $V\leq V_{\text{ref}}$ }} &,&  V \leq V_{\text{ref}}.
    \end{array}
    \right. 
    \label{eq:weight_function}
\end{equation}
The final likelihood is a product of the partial likelihoods in each bin. With this new weight function, bins brighter than $V_{\text{ref}}$ appear in the final likelihood expression multiple times, so it is as if there is the same number of bins brighter and dimmer than $V_{\text{ref}}$. Taking $V_{\text{ref}}$ fainter than the MSTO works most of the time. The entire process of using \texttt{SIESTA} code, and the detailed explanation for the aforementioned likelihood functions is outlined in \cite{Bernardo+2024}. 

The obtained results for the SB clusters are shown, featuring corner plots showcasing the marginalized posteriors (see supplementary Figs. A.1 and B.1 as mentioned in section \ref{cornerplots}). Additionally, an example of the synthetic population corresponding to the best fit is compared with the original data. In each case, it is evident that the solution obtained provides a compelling match to the observed population. The results from visually fitting isochrones serve as priors for the \texttt{SIESTA} code. Further metallicity priors are given based on the spectroscopically derived values from the literature; see Table \ref{tabcomp} for the adopted references.
The so-derived astrophysical parameters for all the clusters are given in Table \ref{tabres}.

To further validate the consistency of our results using PARSEC isochrone models at other wavelengths, we plotted the respective isochrones in multi-band photometry from Gaia and SMASH. This method helps to better constrain the metallicity based on the slope of the RGB and the location of the red clump stars. The catalogs from Gaia and SMASH surveys were cross-matched with our VISCACHA survey data. The membership probabilities of the stars in the decontaminated VISCACHA samples were assigned to the corresponding matched stars in each cluster across all the surveys. CMDs were generated for each survey using their respective filters, and we over-plotted the PARSEC isochrones (in the corresponding filters) using the best-fit parameters derived from the VISCACHA CMDs. 
The best-fit plots for all clusters, including the multi-band CMDs from SMASH and Gaia, are shown in the Appendix (Figs. \ref{MBfit01}, \ref{MBfit02}, and \ref{MBfit03}). In these figures, gray points indicate field stars, while the most probable cluster populations are color-coded based on the membership probability. The solid lines represent the best-fit isochrones obtained by \texttt{SIESTA} on the VISCACHA CMDs.
Notably, the isochrone fitting to the VISCACHA CMD demonstrates strong agreement with data from other surveys, despite their differing filters and wavelength ranges. This multi-band analysis adds significant robustness to our results. 

\section{Astrophysical parameters}\label{sec4}
The astrophysical parameters of our sample clusters were derived homogeneously using the \texttt{SIESTA} code, as described in Section \ref{siesta}, by fitting PARSEC+COLIBRI isochrones to the decontaminated VISCACHA CMDs. We focus primarily on ages and metallicities, as these are the key quantities for constructing the AMR. All derived parameters—including distance, reddening, and binary fraction (BF)—are listed in Table \ref{tabres}, with literature values compiled in Table \ref{tabcomp} for comparison.

Our sample comprises 22 SB clusters representing approximately 67\% of the total 33 clusters cataloged by \citet[][Type = C ]{Bica+20}. 
Of these, 8 clusters have already been analyzed by the VISCACHA team using the Statistical Inference of physical paRameters of sIngle and mUltiple populations in Stellar clusters (SIRIUS) code \citep{2020Souza} and Calcium Triplet (CaT) spectroscopy (see \citetalias{PapIX}). 
In this work, we expand the sample by adding 12 more clusters also observed by VISCACHA, and re-analyze the 8 clusters which are part of the SB region from \citetalias{PapIX}, using the \texttt{SIESTA} code to ensure consistency across the sample clusters. And finally adding HW\,79 and NGC\,339 that are not observed by the VISCACHA survey, building a well defined sample of SB clusters.
\subsection{Homogeneity and comparison with literature}
Several prior studies derived cluster parameters using heterogeneous methods, including fixed values for metallicity and distance, morphological indices, or shallow photometry (see, Table \ref{tabcomp}). Such approaches can introduce systematic biases, particularly when used for AMR construction. Our use of uniform, deep VISCACHA photometry and a consistent Bayesian fitting framework (SIESTA) minimizes these effects, offering more robust results.
The \texttt{SIESTA}-derived parameters show good agreement with spectroscopic metallicities and with previous VISCACHA results (\citetalias{PapIX}), validating the reliability of the method. Differences with some literature values are attributed to differing assumptions in isochrone models, fixed parameters, or photometric depth and quality.
A detailed comparison of results from these literature with the homogeneously derived astrophysical parameters using \texttt{SIESTA} of each cluster is as follows.

\begin{table*}
    \centering
    \caption{Astrophysical parameters derived from the isochrone fits.}
    \label{tabres}

    \begin{tabular}{p{3cm}p{2.5cm}p{2.5cm}p{2.5cm}p{2.5cm}p{3cm}}
         \hline
         \noalign{\smallskip}
         Cluster Name & Age & [Fe/H] & Distance  & ${\text{E(B-V)}}$  & Binary Fraction  \\
         & [Gyr] &&[kpc]& [mag] & -\\ 
        \noalign{\smallskip}
         \hline\hline
        \noalign{\smallskip}
         
         B91      &$0.87\pm0.06$      &$-0.73\pm0.06$       & $57.2\pm0.5$       & $0.07\pm0.01$      & $0.51\pm0.07$     \\ \noalign{\smallskip}
         B98ne    &$3.97\pm0.22$      &$-0.93\pm0.06$       & $54.4\pm1.3$       & $0.07\pm0.01$      & $0.14\pm0.11$     \\ \noalign{\smallskip}
         BS75$^*$ &$2.49\pm0.11$      &$-0.52\pm0.11$       & $62.6\pm2.4$       & $0.03\pm0.02$      & $0.19\pm0.12$     \\ \noalign{\smallskip}
         BS80     &$3.72\pm0.29$      &$-0.95\pm0.08$       & $57.5\pm2.2$       & $0.08\pm0.01$      & $0.15\pm0.09$     \\ \noalign{\smallskip}
         HW20$^*$ &$1.29\pm0.11$      &$-0.78\pm0.05$       & $59.4\pm1.9$       & $0.09\pm0.01$      & $0.14\pm0.07$     \\ \noalign{\smallskip}
         HW27     &$0.58\pm0.08$      &$-0.62\pm0.22$       & $47.2\pm3.8$       & $0.15\pm0.03$      & $0.10\pm0.05$     \\ \noalign{\smallskip}
         HW31     &$3.02\pm0.11$      &$-0.89\pm0.04$       & $55.6\pm1.2$       & $0.11\pm0.01$      & $0.47\pm0.22$     \\ \noalign{\smallskip}
         HW36$^*$ &$2.61\pm0.09$      &$-0.79\pm0.07$       & $62.4\pm2.3$       & $0.01\pm0.02$      & $0.14\pm0.09$     \\ \noalign{\smallskip}
         HW38     &$6.21\pm0.46$      &$-1.20\pm0.17$       & $51.0\pm1.6$       & $0.06\pm0.01$      & $0.24\pm0.14$     \\ \noalign{\smallskip}
         HW44$^*$ &$0.57\pm0.08$      &$-0.94\pm0.09$       & $58.9\pm4.1$       & $0.03\pm0.01$      & $0.27\pm0.11$     \\ \noalign{\smallskip}
         HW47     &$2.11\pm0.16$      &$-0.87\pm0.07$       & $55.5\pm2.6$       & $0.08\pm0.03$      & $0.15\pm0.10$     \\ \noalign{\smallskip}
         HW66     &$4.14\pm0.16$      &$-1.09\pm0.05$       & $59.0\pm1.2$       & $0.16\pm0.01$      & $0.15\pm0.09$     \\ \noalign{\smallskip}
         K37      &$1.58\pm0.01$      &$-0.71\pm0.01$       & $68.5\pm0.7$       & $0.01\pm0.00$      & $0.59\pm0.11$     \\ \noalign{\smallskip}
         K44      &$3.53\pm0.11$      &$-0.78\pm0.06$       & $69.6\pm0.4$       & $0.05\pm0.01$      & $0.68\pm0.05$     \\ \noalign{\smallskip}
         L106     &$1.53\pm0.02$      &$-0.81\pm0.05$       & $49.8\pm1.3$       & $0.07\pm0.01$      & $0.33\pm0.12$     \\ \noalign{\smallskip}
         L112     &$5.81\pm0.57$      &$-1.07\pm0.07$       & $56.9\pm2.9$       & $0.11\pm0.02$      & $0.14\pm0.08$     \\ \noalign{\smallskip}
         L116     &$2.00\pm0.18$      &$-0.86\pm0.05$       & $45.1\pm0.9$       & $0.11\pm0.03$      & $0.23\pm0.18$     \\ \noalign{\smallskip}
         NGC643   &$1.88\pm0.07$      &$-0.65\pm0.07$       & $50.9\pm1.9$       & $0.08\pm0.01$      & $0.94\pm0.06$     \\ \noalign{\smallskip}
         RZ107    &$3.03\pm0.07$      &$-0.95\pm0.04$       & $69.7\pm1.3$       & $0.04\pm0.01$      & $0.20\pm0.10$     \\ \noalign{\smallskip}
         RZ158    &$5.09\pm0.25$      &$-1.13\pm0.05$       & $54.6\pm1.2$       & $0.09\pm0.01$      & $0.11\pm0.05$     \\ \noalign{\smallskip}
            \hline  \noalign{\smallskip}
            \multicolumn{6}{c}{Samples takes from literature}\\       
            \hline  \noalign{\smallskip}
         HW79     &$4.90\pm0.20$      &$-1.26\pm0.03$       & $56.5\pm1.1$       & $0.08\pm0.02$      & $--$     \\ \noalign{\smallskip}
         NGC339   &$6.00\pm0.50$      &$-1.24\pm0.01$       & $56.2$       & $0.06$      & $--$     \\ \noalign{\smallskip}
         \hline
    \end{tabular}
    \begin{tablenotes}
    \item[]$^*$Clusters that are classified as part of the SMC main body based on their relative 3D positions from the SMC center. 
    \end{tablenotes}
\end{table*}
B\,91 was studied by \citet{Glatt2010} and \citet{Rafelski_Zaritsky2005} using shallow photometry from the MCPS, fixed DM, and metallicity, deriving an age of approximately 0.8 Gyr. Our best-fitting isochrone yields parameters of $\text{age} = 0.87\pm0.06$~Gyr, together with $[\text{Fe/H}] = -0.73\pm0.06~{\text{dex}}$.

B98ne has a metallicity estimate derived from the CaT spectroscopy by \citetalias{PapIX}, which is reported to be $-0.96\pm0.05$~dex. This metallicity estimate is consistent with the metallicity of $-0.94\pm0.06$~dex derived from our best fitting isochrone. Additionally, the age of $3.97\pm0.22$~Gyr for B98ne that was derived together with the aforementioned metallicity is in agreement with the age estimate of $3.5\pm0.5$~Gyr reported by \citetalias{PapIX}.

BS\,75 was analyzed by \cite{Glatt2010}, who determined an age of 1.25~Gyr. \cite{Perren2017} derived the values for $\text{age} = 2.5\pm0.10$~Gyr and $[\text{Fe/H}] = -0.48\pm0.26$~dex. Our best fitting isochrone yields $\text{age} = 2.51\pm0.10$~Gyr and $[\text{Fe/H}] = -0.52\pm0.11$~dex in good agreement with the aforementioned results from ASteCA. \citetalias{PapIX} uses the SIRIUS code (\citealp{2020Souza}) deriving the cluster parameters, finding $\text{age} = 2.40\pm0.20$~Gyr, and a metal poor value of $[\text{Fe/H}] = -0.84\pm0.11$~dex. This discrepancy in metallicity may be due to the lack of RGB population in the decontaminated CMD of the cluster affecting the determination of $[\text{Fe/H}]$, as the color and slope of the RGB are very sensitive to the metallicity. 

BS\,80 was analyzed by \cite{Perren2017}, who derived an age of 3.98~Gyr and a metallicity of $-0.88\pm0.65$~dex (note the large uncertainty in the metallicity estimation). Recently, \citetalias{PapIX} analyzed BS\,80 both photometrically and spectroscopically yielding a slightly younger age of $3.0\pm0.5$~Gyr, and CaT metallicity of $-1.00\pm0.08$~dex. Our estimate of $\text{age} = 3.72\pm0.3$~Gyr lies in the middle of the aforementioned values, and the metallicity estimate of $-0.95\pm0.08$~dex is in excellent agreement with the CaT metallicity.

\begin{figure*}[t]
    \centering
    \begin{subfigure}[b]{\textwidth}
         \includegraphics[width=0.49\textwidth,height = 0.45\textwidth]{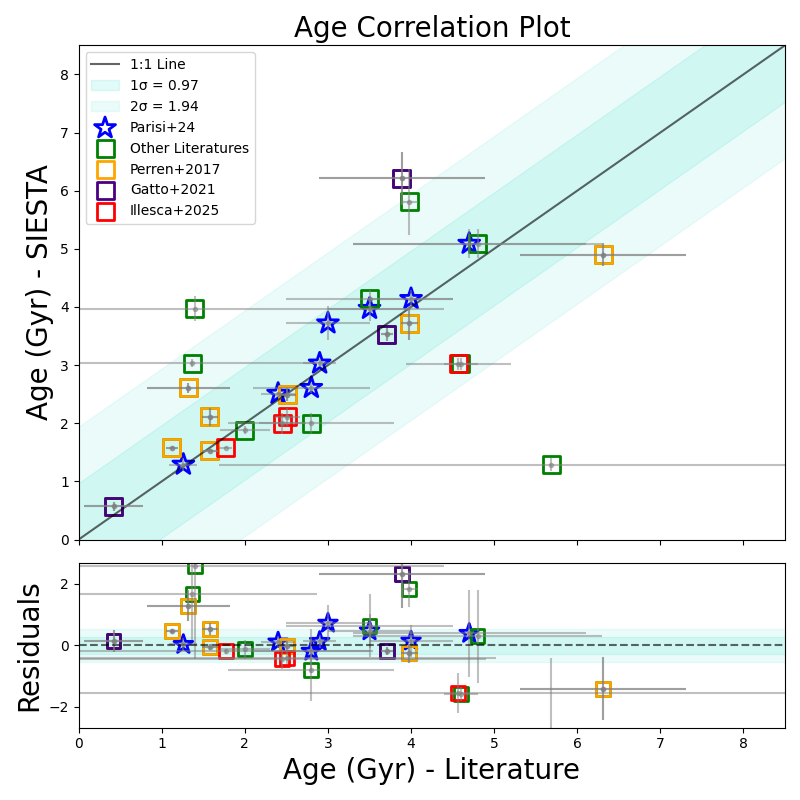}
         \includegraphics[width=0.49\textwidth,height = 0.45\textwidth]{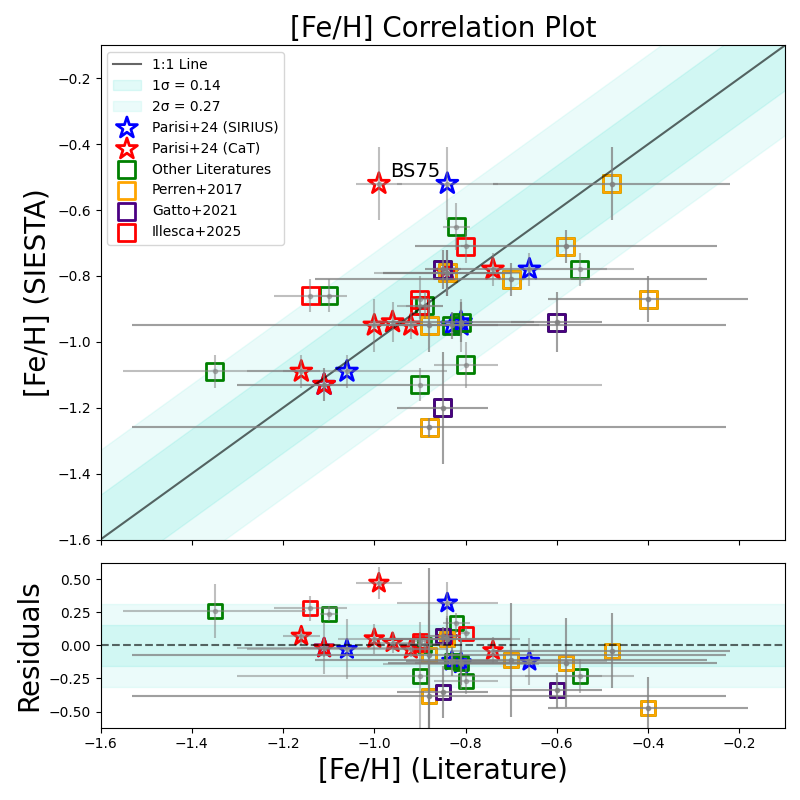}
     \end{subfigure}
    \caption{The correlation plots between the ages (left) and metallicities (right) derived using the \texttt{SIESTA} code is compared with the values from literature excluding previous studies using the VISCACHA data, where the ages and metallicities were derived heterogeneously using different techniques (squares). The blue stars are the ones homogeneously derived by \citetalias{PapIX} using the SIRIUS code, which show good agreement with our estimates. The corresponding residual plots are displayed beneath each main plot. The 1 and 2 sigma regions are shaded to provide a visual representation of the deviations in the correlation plot. Each cluster analyzed in this work has multiple points in the plot from the different literature sources for comparison.} 
    \label{corrSB}
\end{figure*}

HW\,20 was previously classified as an old cluster (5.69 Gyr; \citealp{Rafelski_Zaritsky2005}), revised by \citet{PapI}, using VISCACHA data to $1.10\pm0.10$~Gyr, and ($[\text{Fe/H}] = -0.55\pm0.11)$~dex. Our fits yields $\text{age} = 1.29\pm0.11~{\text{Gyr}}$ and $[\text{Fe/H}] = -0.78\pm0.05$~dex that aligns well with the CaT metallicity from \citetalias{PapIX} ($-0.74\pm0.09$~dex).

HW\,27 (\citealp{1974HW}) stands out as the youngest cluster in our sample. It was previously estimated at 100–200 Myr (\cite{Rafelski_Zaritsky2005,Glatt2010}) based on shallower photometry from the Magellanic Clouds Photometric Survey (MCPS). Both of the aforementioned studies used a fixed distance modulus (DM) and metallicity $Z=0.004$ (\([\text{Fe/H}]\sim -0.6\)~dex) for the SMC clusters to derive the parameters. Our deeper data yields a revised age of $0.58\pm0.08$~Gyr and $[\text{Fe/H}] = -0.62\pm0.22$~dex. 

HW\,31 analyzed by \cite{Rafelski_Zaritsky2005}. They derived an age of $\sim 1.26$~Gyr. \cite{Perren2017} derived the values as $\text{age} = 3.16\pm0.05$~Gyr and $[\text{Fe/H}] = -0.70\pm0.43$~dex. An age of $4.6\pm0.03$~Gyr is derived by \citealp{Parisi2014}, and \cite{deBortili2022} derived the metallicity if the cluster from CaT lines, yielding a value of $-0.89\pm0.04$~dex. The best fitting isochrone from our study yielded a value of $[\text{Fe/H}] = -0.89\pm0.04$~dex, along with an $\text{age} = 3.02\pm0.11$~Gyr.

HW\,36 was recently analyzed both photometrically and spectroscopically by \citetalias{PapIX} deriving its $\text{age} = 2.8\pm0.7$~Gyr and $[\text{Fe/H}] = -0.84\pm0.14$~dex, which are in good agreement with our best-fitting isochrone.
Previously, \cite{Rafelski_Zaritsky2005} and \cite{Gatto2021} estimated an younger age of $\sim1.35$~Gyr, but with fixed DM and metallicity values.

HW\,38 exhibits the largest age discrepancy. Previously estimated to be around 600~Myr (\citealp{Rafelski_Zaritsky2005,Glatt2010}), and an age of $\sim4$~Gyr was derived by \citealp{Gatto2021} using the ASteCA, with fixed distance and metallicity. Our deep photometry and robust analysis yield $\text{age} = 6.21\pm0.46$~Gyr and $[\text{Fe/H}] = -1.20\pm0.17$~dex, placing it among the oldest SB clusters.

HW\,44 is one of the youngest and most metal-poor clusters in our sample. \texttt{SIESTA} gives $\text{age} = 0.57\pm0.08~{\text{Gyr}}$ and $[\text{Fe/H}] = -0.94\pm0.09~{\text{dex}}$. Prior estimation for the cluster parameter was done by \cite{Gatto2021} using the Automated STEllar Cluster Analysis (ASteCA; \citealp{2015Asteca}) code. They derived an age of $0.42\pm0.35$~Gyr, using a fixed DM of 18.98 mag (62.52 kpc) to all the SMC star clusters, and metallicity obtained using the AMR from \cite{Parisi2015}. 

HW\,47 was studied by \cite{Perren2017} and the best fitting isochrone from ASteCA gives values of $\text{age} = 1.58\pm0.10$~Gyr and $[\text{Fe/H}] = -0.40\pm0.22$~dex. The metallicity estimated by \cite{Perren2017} did not agree with the spectroscopically derived value of $[\text{Fe/H}] = -0.92\pm0.04$~dex (\citealp{Parisi2009}). On the other-hand, our best fitting isochrone gives a metallicity of $[\text{Fe/H}] = -0.87\pm0.07$~dex, which is in good agreement with the findings of \cite{Parisi2009} from CaT lines and a self-consistently derived $\text{age} = 2.11\pm0.16 $~Gyr.

HW\,66 has recent estimates from \citetalias{PapIX}, yielding the values of $\text{age} = 4.0\pm0.5$~Gyr and $[\text{Fe/H}] = -1.06\pm0.22$~dex from isochrone fitting using SIRIUS code. They also derived the metallicity from CaT spectroscopy, obtaining a value of $-1.16\pm0.04$ dex. Our best fitting isochrone gives the parameters of $\text{age} = 4.14\pm0.16$~Gyr and $[\text{Fe/H}] = -1.09\pm0.05$~dex, showing exceptional agreement with the previous estimates. \citet{Perren2017} shows a younger age and a more metal-rich estimate for $[\text{Fe/H}]$, a trend that we see with their parameters (see, Fig.\ref{corrSB}). 

K\,37 (L\,58): \texttt{SIESTA} returns \(\text{age}= 1.58 \pm 0.01\)~Gyr and \([ \text{Fe/H}]= -0.78 \pm 0.05\)~dex in excellent agreement with prior spectroscopic and photometric studies, including Maia et al. (2019) and Parisi et al. (2015).

K\,44 analyzed by \cite{Gatto2021} yielded the values for $\text{age} = 3.71\pm0.07$~Gyr and metallicity approximately $[\text{Fe/H}] = -0.85$~dex. \cite{parisi+22} derived a spectroscopic metallicity value of $[\text{Fe/H}] = -0.78\pm0.03$~dex. \cite{Rafelski_Zaritsky2005} derived an older age of approximately 5.6~Gyr. Our best fitting isochrone reproduces the spectroscopic metallicity of $[\text{Fe/H}] = -0.78\pm0.06$~dex. The self consistent parameters along with a metallicity of $-0.78$~dex yields an $\text{age} = 3.53\pm0.11$~Gyr, which is in good agreement with the values derived using ASteCA by \cite{Gatto2021}. 

L\,106 (\citealp{1962Lindsay}) : Our derived \([\text{Fe/H}]= -0.81\pm0.05\)~dex aligns with the CaT value from \cite{Parisi2009}, while our age of $1.53 \pm 0.02$~Gyr improves on earlier estimates with reduced uncertainty.

L\,112 was analyzed by \cite{Parisi2014} deriving an age of $5.10\pm0.30$~Gyr. Furthermore, \cite{Parisi2015} estimated the metallicity using spectroscopy yielding $[\text{Fe/H}] = -1.08\pm0.07$~dex, which is consistent with our estimate. Our age estimate of $5.81\pm0.57$~Gyr agrees with the previous estimates. \cite{Perren2017} using ASteCA found a younger age of $\sim4$~Gyr, and metal-rich estimate of $[\text{Fe/H}] = -0.88\pm0.65$~dex with large uncertainties. 

L\,116 was recently analyzed spectroscopically by \cite{parisi+22}, who used CaT lines to derive a metallicity of $[\text{Fe/H}] = -0.89\pm0.02$~dex. The best fitting isochrone from \texttt{SIESTA} yields a metallicity of $-0.86\pm0.05$~dex, which is in good agreement with the spectroscopic results. \citet{Piatti2001} derived an age of $2.8\pm1.0$~Gyr and a metallicity of $[\text{Fe/H}] = -1.10 \pm 0.20$~dex by using Washington photometry. However, the age estimate carries large uncertainties due to challenges in accurately determining the cluster's MSTO, primarily caused by intrinsic dispersion and photometric errors at faint magnitudes.
Along with metallicity, we homogeneously derived an $\text{age} = 2.00\pm0.18$~Gyr and $\text{D} = 45.05\pm0.91$~kpc. Notably, our distance estimate is slightly higher but consistent with the value reported by \citet{Piatti2025}, who used reddening-corrected CMDs from SMASH and the ASteCA code to determine the cluster’s astrophysical parameters. Their analysis resulted in a slightly older and more metal-poor estimates compared to our findings.

NGC\,643 (L\,111): A complex case due to a crowded Red Clump (RC). 
Our best fitting isochrone yielded the values for $\text{age} = 1.88\pm0.07$~Gyr and $[\text{Fe/H}] = -0.65\pm0.07$~dex. Over three decades ago \cite{Bica+1986} derived the age of the cluster as $1.5\pm0.3$~Gyr, but from integrated photometry. 
\cite{2002AhumadA} derived the age of the cluster as $1.00\pm0.20$~Gyr using integrated spectra. A metallicity of $-0.82\pm0.03$~dex was derived spectroscopically by \cite{Parisi2009}, which is more metal poor than what we found. \cite{Parisi2014} derived an age of $2.0\pm0.3$~Gyr which agrees well with our estimation. 

RZ\,107 yields an $\text{age} = 3.03\pm0.07$~Gyr and $[\text{Fe/H}] = -0.95\pm0.04$~dex from our best fitting isochrone, which is consistent with the results from \citetalias{PapIX}(see; Table \ref{tabcomp}). The metallicity derived from CaT is $-0.92\pm0.04$~dex. 

RZ\,158 was identified as an old cluster (7.3~Gyr) by \citet{Rafelski_Zaritsky2005}. \citet{Bica2022}, using the SIRIUS code on a decontaminated VISCACHA CMD, derived a younger age of $4.8 \pm 1.5$~Gyr and a metallicity of $-0.90 \pm 0.40$~dex. More recently, \citetalias{PapIX} determined a CaT metallicity of $[\text{Fe/H}] = -1.11 \pm 0.05$~dex and $\text{age} = 4.7 \pm 1.4$~Gyr from fitting isochrones. Our homogeneously derived values of $\text{age} = 5.09 \pm 0.25$~Gyr show reasonable agreement with these previous estimates. Additionally, our metallicity estimate of $[\text{Fe/H}] = -1.13 \pm 0.05$~dex exhibits excellent correlation with the CaT metallicity value.

Below we discuss the SB clusters which are not in our sample, but have ages and metallicities derived with good precision.\\

HW\,79 was previously analyzed by the VISCACHA survey \citepalias{PapIX} team using the SIRIUS code (\citealp{2020Souza}) deriving the values for $\text{age} = 4.9\pm0.24$~Gyr and from CaT lines an $[\text{Fe/H}] = -1.26\pm0.03$~dex. The analysis was carried out using the SMASH survey (\citealp{nidever+17}) data due to some technical problems that occurred during the observations with SAMI.

NGC\,339 has been analyzed by \cite{Glatt2008b}, visually fitting isochrones to the CMD obtained from Hubble Space Telescope (HST) photometry and using a fixed value of $[\text{Fe/H}] = -1.19\pm0.10$~dex (\cite{Dacosta1998}). By fitting Padova (\citealp{2000Girardi}) isochrones they derived an age of $6.3$~Gyr. They also used Teramo (\citealp{2004Teramoisoc}) isochrones which yielded an $\text{age} = 6.6$~Gyr.  
\citealp{Narloch2021} derived the values by fitting PARSEC isochrones and found an $\text{age} = 5.6$~Gyr and $[\text{Fe/H}] = -1.10\pm0.12$~dex. 
\cite{Mucciarelli2023} used high resolution spectroscopy estimating the metallicity as $[\text{Fe/H}]=-1.24\pm0.01$~dex, which is the one we adapted for our AMR. And we used an average value of $\text{age} = 6.0\pm0.5$~Gyr for the AMR. 
\subsection{Statistical comparisons}
Fig. \ref{corrSB} presents a statistical comparison of SIESTA-derived parameters with those from the literature. The literature values were picked from the compilation shown in Table \ref{tabcomp}, meaning that for each cluster there are multiple values from the literature. In particular, we highlight sources that have clusters analyzed in a homogeneous way: \citet{Perren2017, Gatto2021, Piatti2025}, besides \citetalias{PapVII} from our collaboration.
\citet{Perren2017} derived the parameters with PARSEC v1.1 theoretical isochrones (\citealp{Bressen2012}) and a log-normal IMF by \citet{Chabrier2001}, using the ASteCA code, whereas we have used PARSEC+COLOBRI isochrones (v3.7) and the \citet{kroupa2001} IMF. It is also to be noted that ASteCA used a Poisson likelihood rate (\citealp{Dolphin2002}) to compare the synthetic and observed CMD. 
\citet{Gatto2021} fixed the distance to the SMC and metallicities to derive the ages, along with their structural parameters of the cluster. 
\citet{Piatti2025} used the reddening corrected magnitudes to derive the age, metallicity, distance and mass of the clusters using the ASteCA, with PARSEC v1.2S isochrones (\citealp{Bressen2012}), and the \citet{Kroupa2002} IMF. 

The left panel of the figure \ref{corrSB} presents a correlation plot comparing the cluster ages derived using the SIESTA method with values from various literature sources. The black solid line represents the 1:1 correlation, indicating good agreement. 
The shaded regions represent the $1~\sigma$ and $2~\sigma$ confidence intervals, and the residual plot below highlights the deviations, showing that most differences remain within the $2~\sigma$ range, although a few outliers exist. 
The ages derived by \citetalias{PapIX} (blue stars) show strong agreement with the \texttt{SIESTA} values, closely following the 1:1 line. Most points have small residuals, indicating minimal deviation from literature values.
The \citet{Perren2017} (yellow squares) generally align well with \texttt{SIESTA} results but exhibit some dispersion, especially at intermediate $1-2$~Gyr) and older $>6$~Gyr) ages.
The \citet{Gatto2021} values (purple squares) remain mostly (2 out of 3) consistent with \texttt{SIESTA}, with a one outlier (HW\,38) with an underestimated age compared to \texttt{SIESTA}. 
The red squares correspond to the estimates from \citet{Piatti2025}, which appear to favor slightly older ages compared to \texttt{SIESTA}.
The heterogeneous estimates from other literature (green squares) display more scatter, with some clusters deviating significantly from the 1:1 line, particularly at higher ages ($>4$~Gyr). This suggests the systematic differences in the methods used to determine ages in different studies. Furthermore, variations in survey characteristics (e.g., completeness and depth) may also impact the determination of astrophysical parameters of the star clusters.
\begin{figure}[t]
    \centering
      \begin{subfigure}[b]{\textwidth}
         \includegraphics[width=0.5\textwidth,height = 6cm]{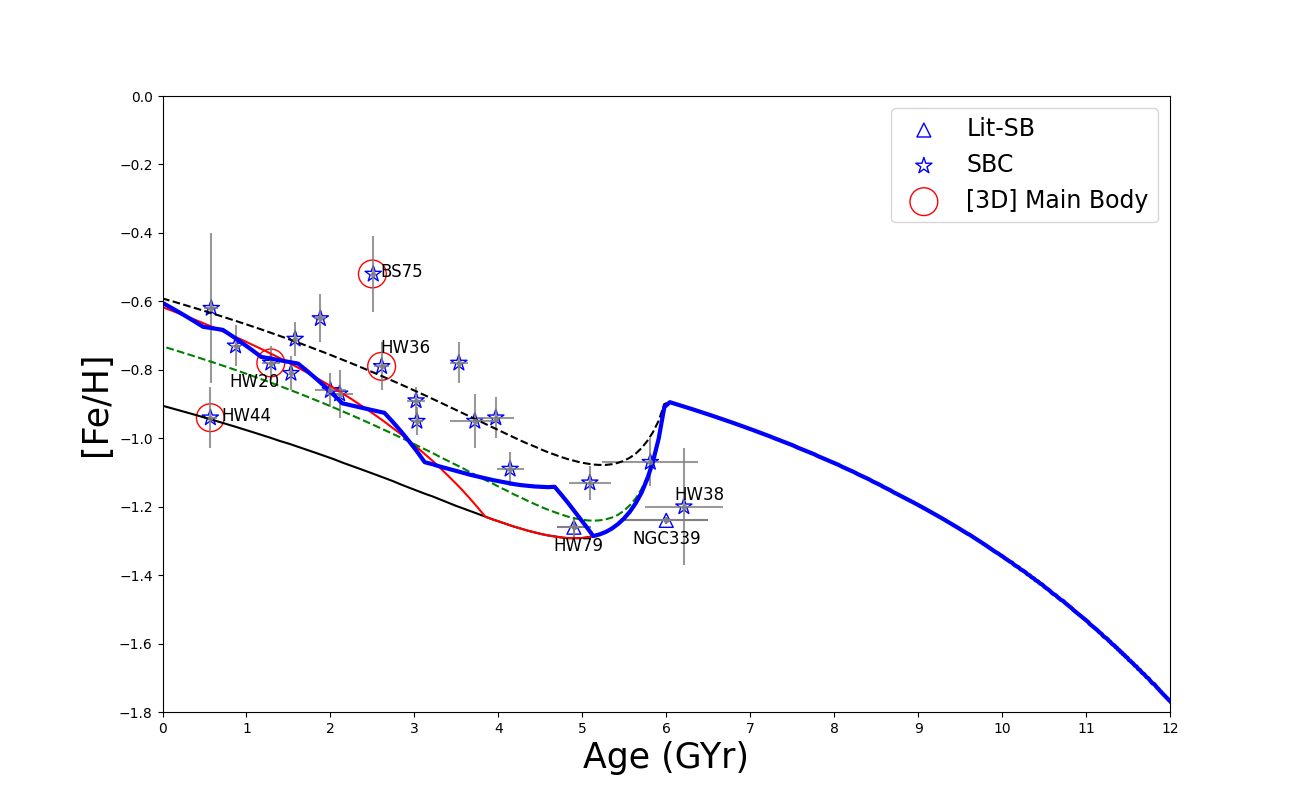}
     \end{subfigure}
    \caption{AMR of the SMC SB clusters. 
    Triangles represent results from an extended sample from the literature.
    Clusters classified as part of the main body based on the 3D classification are labeled and highlighted with red circles.
  The green curve is the model for the WH clusters \citepalias{PapVIII}, representing the major merger 6~Gyr ago resulting in an enhanced SFR. The red curve is similar, but the enhancement in SFR is delayed for 2~Gyr after the major merger. The blue solid line is the major merger model followed by a sequence of bursts, and this is our best fitting model for the WH clusters from \citetalias{PapVIII}. The black thick line depicts the scenario without enhanced star formation after the major merger, whereas the black dotted line is the merger with a mass ratio 1:2.} 
    \label{Imgamr}
\end{figure}

The right panel (Fig. \ref{corrSB}) displays the correlation between metallicities obtained with \texttt{SIESTA} and those from the literature. As in the age plot, the black solid line represents the 1:1 correlation. 
The \citetalias{PapIX} values derived from SIRIUS (blue stars) and CaT (red stars) show a strong correlation with \texttt{SIESTA}, as expected because we use the CaT metallicities as a prior whenever available. 
The SB cluster outside the $2~\sigma$ region is BS\,75, for which we derived a higher metallicity compared to the estimate of $-0.99\pm0.10$~dex from CaT lines (\citetalias{PapIX}). However, it is worth noticing that the parameters obtained from isochrone fitting using the SIRIUS code on the VISCACHA CMD yield $[\text{Fe/H}] = -0.84\pm0.11$~dex, which is again metal rich compared to the CaT value.
And for NGC\,643 our estimate of $[\text{Fe/H}] = -0.65$~dex deviates from the slightly more metal-poor spectroscopic value of -0.82, maybe due to the crowded RC region. 
The \citet{Perren2017} points mostly (5 out of 7) align with \texttt{SIESTA}, and they lie just below the 1:1 line, suggesting a small underestimation relative to \texttt{SIESTA} metallicities ($\sim0.1$~dex). The two extreme outliers from \citet{Perren2017} are HW47 and HW79 for which ASteCA returned metal rich values with large error bars. 
The \citet{Gatto2021} values (purple squares) appear to be slightly more metal-rich compared to \texttt{SIESTA} estimates, but the deviation is generally within $1~\sigma$ region. It is worth noticing that they keep the distance and metallicity (inferred from the AMR) fixed for fitting the isochrone, which potentially leads to degenerated results.
The \citet{Piatti2025} metallicities (red squares) show a small systematic offset, with slightly more metal-poor values compared to \texttt{SIESTA}. 
The values heterogeneously estimated by other literature (green squares) exhibit the largest scatter, with some data points deviating significantly from the 1:1 line. This suggests that different studies have used a wide range of methodologies, possibly involving varying spectral analysis techniques, different photometric datasets, and approximations in SMC distance and reddening. 
 
\citet{Piatti2025} and \citetalias{PapIX} provide a self-consistent analysis, whereas \citet{Piatti2025} employ different data and methodologies; nonetheless, both studies show overall agreement with our results. \citet{Perren2017}) report large uncertainties in their results. \citet{Gatto2021} constrained distance and metallicity during the fitting process, which may introduce degeneracies. Other studies in the literature rely on shallower or lower-quality photometry or older isochrone models.

\section{Interpreting the chemical evolution of the SMC}
\subsection{The age-metallicity relation of the southern SMC based on star clusters}
We present the most up-to-date AMR for the SMC clusters in the SB and WH regions. We have analyzed 67\% of the total SB and 60\% of the total WH clusters, catalogued by \citet{Bica+20}. For clusters not included in our sample, we adopted the best age and metallicity estimates available from the literature (see table \ref{rescomp}). We propose that the combined sample can be treated as quasi-homogeneous, allowing us to trace the most up-to-date AMR of the SB and WH combined, and providing valuable insights into their chemical evolution histories.

\citetalias{PapVIII} derived the AMR for the WH clusters by visually fitting isochrones to the VISCACHA CMDs, which shows excellent agreement with the \texttt{SIESTA} derived parameters using the same data from the current work (detailed in Appendix \ref{WHCs}). The WH clusters exhibit a distinct chemical enhancement during the early stages of the SMC’s evolution, between 10 and 6 Gyr ago, reaching approximately $-0.8$~dex. At the end of this period, a major merger event caused a sharp decline in metallicity by about 0.5~dex. Following this downturn, chemical enrichment resumed, eventually reaching back to $\sim -0.8$~dex around 1 Gyr ago (models on Fig. \ref{Imgamr}).

The SB clusters exhibit a similar trend in the AMR with all clusters younger than the $\sim$6~Gyr metallicity dip (Fig. \ref{Imgamr}). Three of the SB clusters (L112, HW38, and NGC339) have ages close to the metallicity dip. Among them, L112 aligns with the metallicity dip in the AMR and serves as the connection to the AMR trend of the WH. 
Our estimate of 5.81 Gyr, obtained in a self-consistent method, places L112 in a well-defined and reliable position in the AMR.
For HW38 with a broad and dense MSTO, our self-consistently derived parameters improves on earlier estimates with reduced uncertainties, placing the cluster in a well-defined position in the AMR.
NGC339 is not in our sample of VISCACHA clusters, but was extensively studied by \citealp{Glatt2008b,Narloch2021} deriving an average age of 6~Gyr. The metallicity of the cluster is also derived using CaT lines (\citealp{Dacosta1998}) and high-resolution spectroscopy (\citealp{Mucciarelli2023}). 

\begin{figure}[t]
    \centering
      \begin{subfigure}[b]{\textwidth}
         \includegraphics[width=0.5\textwidth,height = 6.2cm]{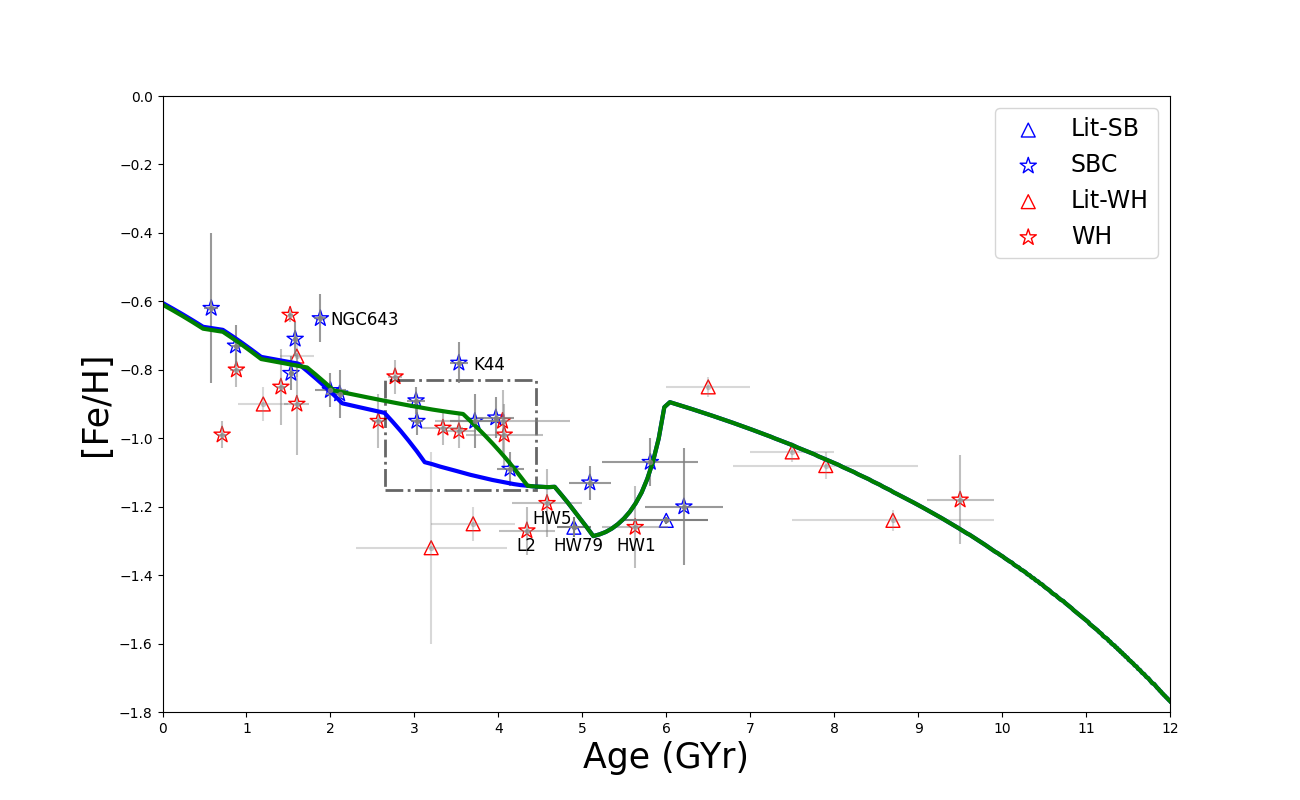}   
     \end{subfigure}
    \caption{Comparison between the best fitted model from \citetalias{PapVIII} with the modified one from the current work. The AMR with all the clusters from SB (excluding the ones classified as [3D] main body as shown in Fig. \ref{Imgamr}) and WH clusters. The best-fitting major merger model from \citetalias{PapVIII}, derived using only the WH clusters, is over plotted as the solid blue line. The green solid line represents our newly modified major merger model, accounting for all the clusters from SB and WH together. 
  This best-fit green solid curve illustrates a merger model with a 1:4 mass ratio 6~Gyr ago, followed by enhanced SFR from multiple bursts as presented in Fig. \ref{SFRbursts}. The region enclosed by the black rectangle highlights the most significant difference between the old (\citetalias{PapVIII}) and new models.  } 
  
    \label{Imgamrnew}
\end{figure}

The SB cluster HW79 along with the WH clusters HW1, HW5, and L2 define the depth of metallicity dip, which is about $\sim0.5$~dex (see Fig. \ref{Imgamrnew}). HW79 has it age derived from the SMASH data using the SIRIUS code and spectroscopically derived metallicity \citepalias{PapIX}, so its position in the AMR is reliable. All other clusters are analyzed in the current work (explained in section \ref{sec4}).

The SB clusters in the age range $\sim$3 to $\sim$4.5~Gyr are more metal-rich than the best-fit models derived using the WH clusters from \citetalias{PapVIII} (see 
Fig \ref{Imgamr}). The new results with \texttt{SIESTA} for the WH clusters also show a similar trend along with the SB clusters around this age range. We have emulated multiple bursts from \cite{2015RUBELE} (around 5~Gyr ago) and \cite{2022Massana} (from 3 to 1~Gyr ago in intervals of 0.5~Gyr). 
The results from our current work portray that there is an additional star-burst between 3 and 4.5~Gyr ago, as shown in Fig. \ref{Imgamrnew} (the region enclosed by the rectangle), and we have adapted the model accordingly; this will be explained in detail in the next section.

The three outliers from SB clusters are NGC643, BS75, and K44; they are discussed in detail in section \ref{sec4}. 
NGC 643 and BS75 exhibited discrepancies in metallicity, with \texttt{SIESTA} derived $[\text{Fe/H}]$ values appearing more metal-rich than those from CaT lines, though BS75's estimate aligns more closely with \citet{Perren2017}. To maintain homogeneity, we adopted \texttt{SIESTA} values for the AMR, but even if we consider the spectroscopic values from \citetalias{PapIX}, they do agree well with the models. 
K44 is a genuine outlier as the ages and metallicities agree very well with the prior estimates, with the derived parameters showing a large fraction of binaries. The young clusters in the WH show a spread in metallicity around 1-2~Gyr, whereas the only SB cluster that lies as an outlier is the young and metal-poor cluster, HW44, having an age of 0.57~Gyr and metallicity $-0.94$~dex. \cite{Gatto2021} derived a metal-rich value of $-0.6$~dex and an age of $0.42\pm0.35$~Gyr. 
Furthermore, the clusters (HW44, HW20, HW36, BS75) encircled as [3D] main body in Fig. \ref{Imgamr} are the ones that are part of the SMC main body based on the 3D classification (for more details on the method refer \citealp{PapIII} and \citetalias{PapIX}).
In summary, even if we consider these clusters in the overall analysis, this will not significantly affect the conclusions on the chemical evolution of the SB and WH regions.

\subsection{Chemical evolution models}
The chemical evolution history of the SMC is very complex, including an intrinsic metallicity dispersion for intermediate-age star clusters, as shown by e.g. \citet{Parisi2009, Parisi2015, PapI, PapIV,parisi+22, milone+22}. Some efforts to disentangle it have been made by many, like \cite{PT98,Tsujimoto&Bekki_2009,Diaz&Bekki2012,nidever+20,Dias+14,Dias+16,parisi+22,deBortili2022,PapVII,PapVIII,PapIX}, and references therein. For example, a solution to the large dispersion observed in metallicity for the SMC star clusters was given by \cite{Dias+16}. They divided the SMC into different sub-regions showing that the analysis of a single region revealed a clear chemical enrichment with low metallicity dispersion. Moreover, \cite{Tsujimoto&Bekki_2009} proposed the major merger model that closely aligns with the empirical findings from the AMR of a sample of SMC star clusters. They suggested that the probable cause of the $\sim0.3$~dex dip in metallicity at 7.5~Gyr is the infall of metal-poor gas having a mass of $\sim 10^8~M_\odot$. In our work, we have adapted this very same model, the framework by \cite{Dias+16} and analyzed a specific region of the SMC with the most up-to-date AMR of its star clusters. 

Following the same framework and looking only at the W/B region of the SMC, \citetalias{PapVII} revealed that this region has a unique chemical enrichment, with younger clusters formed in situ from the gas stripped from the inner disc of the SMC. \cite{deBortili2022} identified two groups of clusters in the SMC main body co-existing. The metal-rich group tends to follow the chemical enrichment pattern similar to that of the field stars by \cite{2004HZ}, whereas the metal-poor group shows no sign of enrichment. 

\citetalias{PapVIII} corroborated the preliminary results from the VISCACHA survey \citep{dias2019} revealing the $\sim 0.5$~dex metallicity dip of the WH clusters about 6 Gyr ago.
They presented a new model for the WH clusters, which proposes that the best plausible explanation for the metallicity dip is a major merger that took place 6~Gyr ago. Fig. \ref{Imgamr} represents these models from \citetalias{PapVIII} specifically for WH clusters, where we used visually fitted isochrones to derive their parameters.
The models in Fig. \ref{Imgamr} are discussed in detail by \citetalias{PapVIII} (refer section 5.2), and \cite{Tsujimoto&Bekki_2009}. Briefly, this study models the AMR of the SMC by incorporating the effects of major mergers. The star formation rate (SFR) is linked to the gas fraction and adjusted for both normal and enhanced scenarios triggered by major mergers or external tidal interactions. The major merger model assumes a 1:4 mass ratio with an increased SFR 6 Gyr ago (green dotted line in Fig. \ref{Imgamr}), which fits the observed dip in [Fe/H]. Alternatively, a model with a more recent SFR enhancement 4 Gyr ago with the star formation rate coefficient, $\nu = 0.08~Gyr^{-1}$ (red solid line in Fig. \ref{Imgamr}). And the solid blue line in Fig. \ref{Imgamr} represents the model adding multiple 
bursts of star formation at the epochs identified by \cite{2015RUBELE} and \cite{2022Massana}. The black solid and dashed curves (see Fig. \ref{Imgamr}) wrap the data in the metal poor and metal rich boundaries, respectively. Black curve illustrate the model with a mass ratio 1:4 with no star formation enrichment after the merger. On the other hand, the black dotted curve is the one with a mass ratio 1:2 followed by enhanced SFR ($\nu = 0.065~Gyr^{-1}$).  

In this work, we focus on the entire southern region of the SMC outskirts and
we propose a unified model for the SMC clusters in the WH and SB regions showing similar trends in the AMR. 
The dispersion in metallicity is still low even if we combine these two regions, and both of them follow the major merger model from \citetalias{PapVIII}. 
In Fig. \ref{Imgamrnew}, we present comparison between the best-fitted model for the WH clusters (blue solid line) and modified model for the combined WH and SB clusters (green solid line, this work).
It is important to note that the best-fitted model for the WH clusters alone (denoted by the blue solid line) follows the joint AMR of WH and SB clusters, except around 3-4 Gyr ago, when the model indicates lower metallicity in comparison to the data, which was fitted better with the new model (green line) by adding a new star formation burst peak at $\sim 4$ Gyr.
The new major merger model assumes a mass ratio of 1:4, with the occurrence of the metal poor merger 6~Gyr ago causing the dip in metallicity. Later, this major merger event triggered an enhancement in SFR, enriching the $[\text{Fe/H}]$ value. Furthermore, this, along with other more recent interactions of the MCs might have generated a series of bursts in SFR (e.g. \citealp{2013Weisz,2013cignoni,2015RUBELE, 2022Massana}). 
\begin{figure}[t]
    \centering
    \includegraphics[width=0.5\textwidth,height = 5cm]{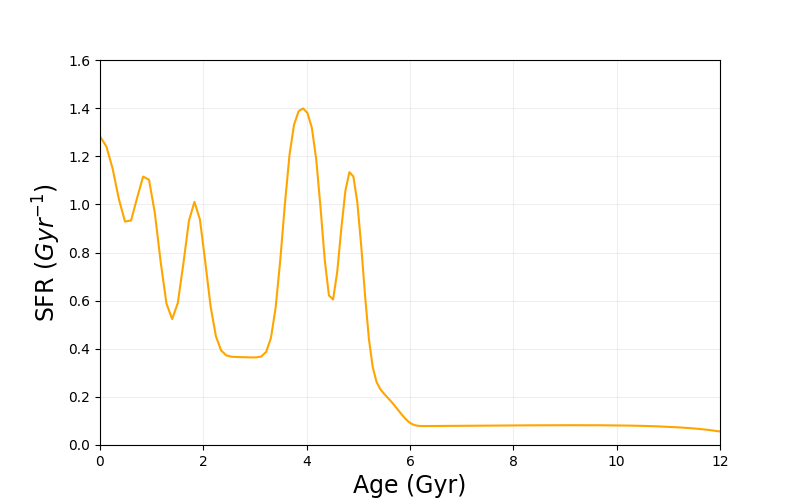}
    \caption{Star formation history of the SMC derived from star clusters.
    The figure illustrates the SFRs corresponding to our best fitted model (green curve in Fig. \ref{Imgamrnew}). The prominent peaks show enhancement in SFR followed by the merger ($3.5-6$~Gyr) and then due to the LMC-SMC interactions ($<3$ Gyr). A Gaussian smoothing with a sigma value of 1.8 Gyr was applied to the model to approximate the broadening of the observed peaks and provide a reasonable visual match.}
    \label{SFRbursts}
\end{figure}

In our best-fitted model (green curve in Fig. \ref{Imgamrnew}) the series of bursts are modelled so as to best reproduce the observed AMR, as illustrated in Fig. \ref{SFRbursts}. The predicted timing and strength of busting modes of star formation, adapted from \cite{2015RUBELE} and \cite{2022Massana}, are loosely constrained to better match the observed AMR. 
From the oldest, the first burst in SFR after the merger is at 5 Gyr.
\cite{2013Weisz} using the Hubble Space Telescope (HST) data (\citealp{2006HSTdata}) derived the star formation history (SFH) of the LMC and SMC. They identified many interesting features in the SFH of the MCs, and the burst at $\sim4.5$~Gyr is one among them that support our findings. \citet{rubele+18} also shows a higher SFR from $4 -6$~Gyr age range. 
The next burst in our model is around 4~Gyr, 
(this part of the AMR is enclosed by the rectangle in the Fig. \ref{Imgamrnew}), a key difference from the models for WH clusters (blue solid line). This peak in SFR is absent in the previous studies focusing on the SMC main body (\citealp{2004HZ,2015RUBELE}), as illustrated in Fig. \ref{SFRbursts}, whereas \citet{rubele+18} illustrates higher SFR compared to \citet{2015RUBELE}. Our SFH, derived from clusters in the SMC outskirts, likely reflects the stronger imprint of the major merger event, whose effects are expected to be more pronounced in the outer regions of the SMC. 
From the observed AMR, the dip in [Fe/H] is already seen around 6~Gyr and subsequently there is an overall [Fe/H] increase until the age of $\sim3$~Gyr. Thus, the prolonged burst suggested by \citet{2015RUBELE,rubele+18} might be compatible with the AMR within an uncertainty in the model prediction.
Followed by this we have a series of bursts from 2.5~Gyr to present, owing to the recent interactions with the MW and LMC, and is supported by the findings of \cite{2022Massana}, who identified multiple bursts in SFR at 3, 2, 1.1, and 0.45~Gyr. Furthermore, \cite{2004HZ} and \cite{2013cignoni} have found bursts of star formation at 1-3~Gyr. \cite{2011Piatti} analyzed star clusters in the SMC older than 1~Gyr, and found the epochs of enhanced cluster formation at $\sim2$~Gyr and $5-6$~Gyr ago. In summary, we point out that the prior works on the SMC SFH are complementary to what we get from star clusters (Fig. \ref{SFRbursts}). 
\cite{2012Piatti} has shown that the AMR of the field stars and star cluster population share the same chemical evolution histories. This latter conclusion adds to the robustness of our results when compared with the findings from different methods involving field stars and star clusters. 

\subsection{Our perspective on the chemical evolution of SMC}

The scenario that best reconciles the AMR of star clusters in the WH and SB regions of the SMC involves a major merger event occurring approximately 6~Gyr ago. This event likely introduced a substantial inflow of pristine, metal-poor gas, leading to a noticeable drop in metallicity. The subsequent epochs were marked by enhanced star formation, as previously discussed and illustrated by the green solid line in Fig. \ref{Imgamrnew}.

We propose that the present-day SMC is the remnant of this merger between two smaller galaxies. The event does not appear to have been extremely violent, as the internal mixing of gas over the last 6 Gyr has been inefficient. This is evident from the distinct AMRs observed in the WH+SB regions (e.g. \citetalias{PapVIII} and this work) compared to the wing and bridge region \citepalias{PapVII}, based on VISCACHA cluster data. The findings are further supported by \citet{Mucciarelli2023}, who used high-resolution spectroscopy of 206 SMC stars across three different fields near globular clusters, revealing regional variations in metallicity, radial velocities, and abundance ratios.

Recent studies reinforce this interpretation. \citet{2024Murray} used high-resolution HI observations to trace the structure of the interstellar medium, unveiling a new model in which the inner 4 degrees of the SMC consist of two superimposed but chemically and kinematically distinct star-forming systems. This supports the notion that inefficient gas mixing has persisted, and that the two components have retained separate identities and evolutionary histories, despite their overlapping spatial distribution.

Such a merger scenario is not improbable. Observational evidence suggests that the Magellanic Clouds are the most massive members of a broader magellanic group of dwarf galaxies, some of which likely survived infall into the MW and became its satellites alongside the LMC and SMC \citep[e.g.][]{sales+2017,kallivayalil+2018,patel+2020}. 
Following the merger, the SMC has experienced multiple interactions with the LMC, triggering recurrent starbursts \citep[e.g.][]{HarrisZaritsky2006,Diaz&Bekki2012,rubele+18,zivick+18,2022Massana}. Notably, over the past $\sim$2 Gyr, as the MCs entered the MW’s gravitational domain, the SFR increased significantly \cite[e.g.][]{nidever+20}.

Our findings derived from the AMR, chemical evolution modeling, and the updated SFH provide strong constraints for understanding the SMC’s chemical evolution. These results also offer critical insights into the dynamical evolution of the magellanic system. To validate this scenario, detailed chemo-dynamical simulations will be essential.

\section{Summary and conclusions}
In this study, we conducted an analysis of the $V$, $V-I$ CMDs of the SB clusters in the SMC, leveraging deep and resolved photometry obtained from the VISCACHA survey. A decontamination method was employed, utilizing statistical comparisons of star samples from the cluster region and an offset field, following the procedures outlined in \cite{Maia2010}. By applying a homogeneous methodology with the \texttt{SIESTA} code (\citealp{Bernardo+2024}), the age, metallicity, distance, and reddening of the clusters were estimated by fitting PARSEC isochrones to the decontaminated CMDs. The fitted isochrones show good correlation with the data from SMASH and Gaia, adding an extra layer of robustness to the results. 
Comparisons with previous studies, including \citetalias{PapIX}, \citet{Perren2017,Gatto2021,Piatti2025},
and other sources from the literature were discussed case by case. The comparison between our homogeneously derived parameters reduced the scatter in age and metallicity as seen from the heterogeneously derived parameters from the literature compilation.
These variations highlight the importance of using a consistent approach to derive cluster parameters, ensuring reliable analysis of the SMC chemical evolution using its star cluster populations.

The chemical evolution history of the SMC is very complex, and possibly the most logical way to disentangle it is by analyzing region by region (\citealp{Dias+16}, \citetalias{PapVII, PapVIII, PapIX}, \citealp{deBortili2022}). In this study, we compiled all the SB and WH clusters observed by the VISCACHA survey, along with data from a few selected literature sources to construct the most up-to-date AMR for these regions.
The observational AMR of these regions shows a low dispersion in metallicity, with a noticeable dip occurring around 6 Gyr ago. Chemical evolution models were run to determine the best fit for the observed AMR trend. The most plausible explanation for this pattern is a major 
merger with a mass ratio of 1:4 and pristine metallicity gas content, which subsequently triggered an enhanced SFR following the event. The AMR was also crucial to identify a new starburst $\sim$4\,Gyr ago, besides the other multiple starburst peaks previously identified.

Our findings provide strong constraints for chemical evolution models of the SMC with implications for the dynamical evolution of the whole magellanic system. Future chemo-dynamical models of the evolution of the LMC, SMC and other MW satellites must be able to explain the observational evidence revealed in this work.
\section*{Data availability}\label{cornerplots}
The SIESTA results for the SB and WH (Figs. A.1 and B.1, respectively) clusters are published as supplementary figures,on Zenodo and are available at: \url{https://doi.org/10.5281/zenodo.15684993}.

\begin{acknowledgements}
SSa and BDi acknowledge support by ANID-FONDECYT iniciación grant No. 11221366. BDi also acknowledges the support by the ANID BASAL project FB210003.
T. T. acknowledges the support by JSPS KAKENHI Grant No. 23H00132.
D.M. gratefully acknowledges the support by the ANID BASAL projects ACE210002 and FB210003 and by Fondecyt Project No. 1220724.
F. Maia acknowledges financial support from Conselho Nacional de Desenvolvimento Científico e Tecnológico - CNPq (proc.404482/2021-0) and from FAPERJ (proc. E-26/201.386/2022 and E-26/211.475/2021).
BPLF acknowledges financial support from Conselho Nacional de Desenvolvimento Científico e Tecnológico (CNPq, Brazil; proc. 140642/2021-8) and Coordenação de Aperfeiçoamento de Pessoal de Nível Superior (CAPES, Brazil; Finance Code 001; proc. 88887.935756/2024-00).
MCP acknowledge support by the Argentinian institutions CONICET (Consejo Nacional de Investigaciones Científicas y Técnicas), SECYT (Universidad Nacional de Córdoba) and ANPCyT (Agencia Nacional de Promoción Científica y Tecnológica).
OKS acknowledges partial financial support by UESC (proc. 073.11157.2022.0033250-38).
D.G. gratefully acknowledges the support provided by Fondecyt regular no. 1220264.D.G. also acknowledges financial support from the Direcci\'on de Investigaci\'on y Desarrollo de la Universidad de La Serena through the Programa de Incentivo a la Investigaci\'on de
Acad\'emicos (PIA-DIDULS).
NSF–DOE Vera C. Rubin Observatory is a Federal project jointly funded by the National Science Foundation (NSF) and the Department of Energy (DOE) Office of Science, with early construction funding received from private donations through the LSST Corporation. The NSF-funded LSST (now Rubin Observatory) Project Office for construction was established as an operating center under the management of the Association of Universities for Research in Astronomy (AURA). The DOE-funded effort to build the Rubin Observatory LSST Camera (LSSTCam) is managed by SLAC National Accelerator Laboratory (SLAC).
APV acknowledges the DGAPA-PAPIIT grant IA103224.\\
Based on observations obtained at the Southern Astrophysical Research (SOAR) telescope (projects SO2016B-018, SO2017B-014, CN2018B-012, SO2019B-019, SO2020B-019, SO2021B-017), which is a joint project of the Ministério da Ciência, Tecnologia, e Inovação (MCTI) da República Federativa do Brasil, the U.S. National Optical Astronomy Observatory (NOAO), the University of North Carolina at Chapel Hill (UNC), and Michigan State University (MSU).
This work has made use of data from the European Space Agency (ESA) mission {\it Gaia} (\url{https://www.cosmos.esa.int/gaia}), processed by the {\it Gaia} Data Processing and Analysis Consortium (DPAC, \url{https://www.cosmos.esa.int/web/gaia/dpac/consortium}). Funding for the DPAC has been provided by national institutions, in particular the institutions participating in the {\it Gaia} Multilateral Agreement.
This research uses services or data provided by the Astro Data Lab at NSF’s NOIRLab. NOIRLab is operated by the Association of Universities for Research in Astronomy (AURA), Inc. under a cooperative agreement with the National Science Foundation.
\end{acknowledgements}

\bibliography{biblio.bib}

\begingroup
\appendix
\onecolumn
\section{Multiband CMDs and Corner Plots} \label{APPB}
The corner plots generated with SIESTA, showing the best-fitting isochrones and derived parameters for the SB clusters, are provided as supplementary material Fig. A1 (see, section~\ref{cornerplots}).
 The corresponding best fitting isochrones of the SB clusters are then plotted over multiband CMDs from Gaia and SMASH (Figures : \ref{MBfit01}, \ref{MBfit02}, \ref{MBfit03}). The literature survey of astrophysical parameters and their respective references are given table \ref{rescomp}. 

\begin{figure*}
    \centering

    \begin{subfigure}[b]{\textwidth}
         \includegraphics[width=0.98\textwidth,height = 3.8cm]{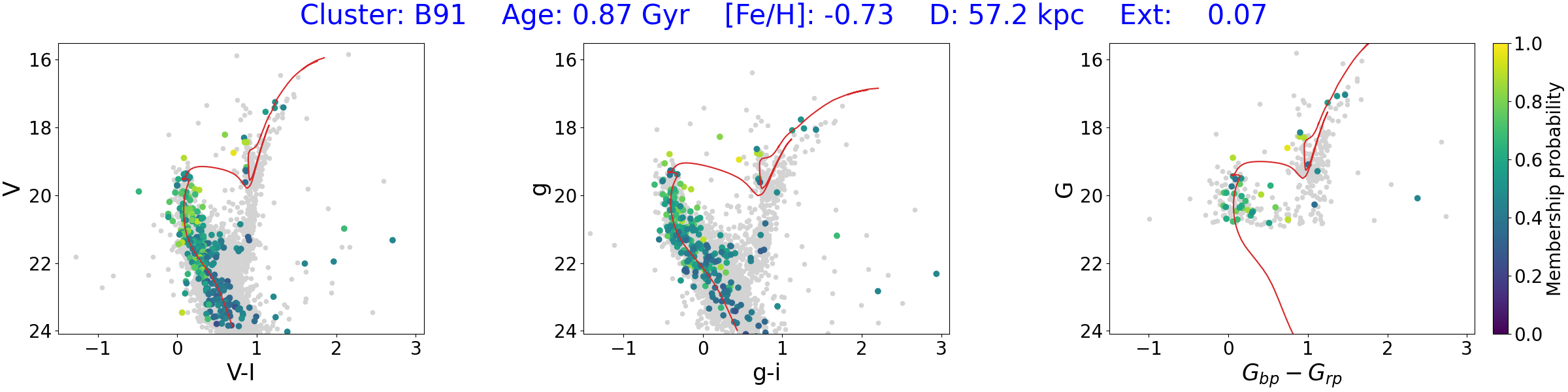}
     \end{subfigure}

      \begin{subfigure}[b]{\textwidth}
         \includegraphics[width=0.98\textwidth,height = 3.8cm]{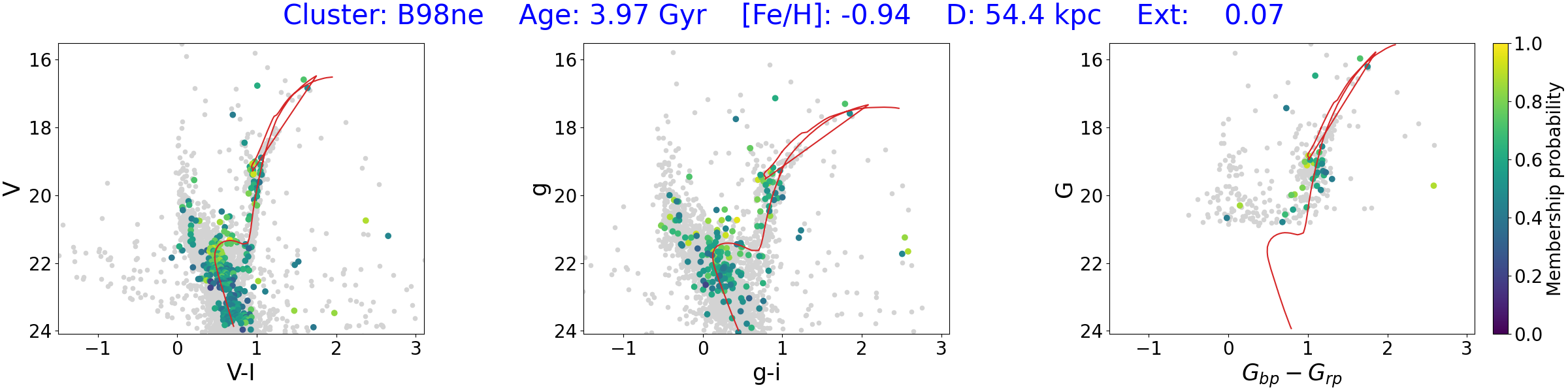}
     \end{subfigure}

     \begin{subfigure}[b]{\textwidth}
         \includegraphics[width=0.98\textwidth,height = 3.8cm]{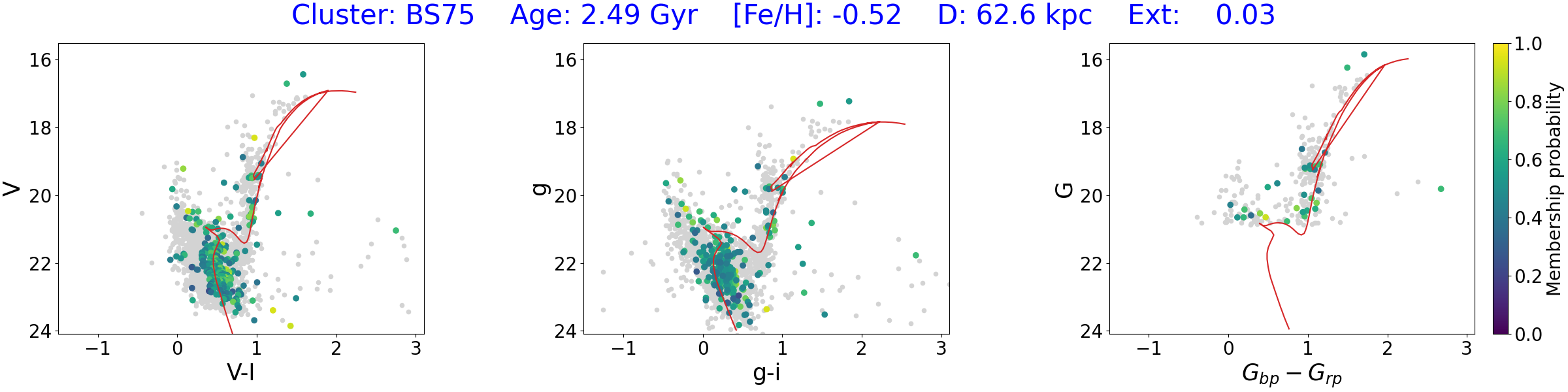}
     \end{subfigure}

      \begin{subfigure}[b]{\textwidth}
         \includegraphics[width=0.98\textwidth,height = 3.8cm]{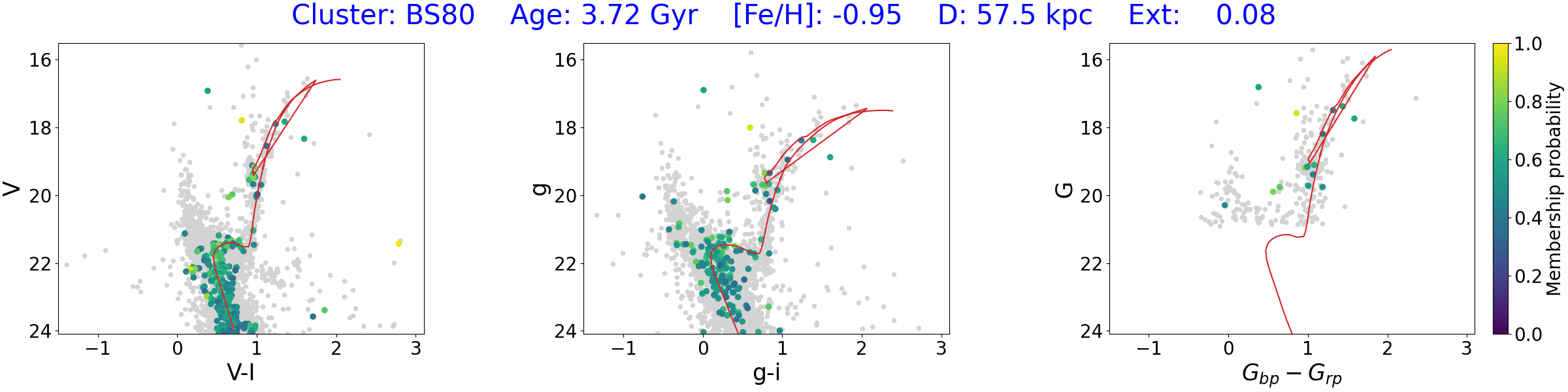}
     \end{subfigure}

    \begin{subfigure}[b]{\textwidth}
        \includegraphics[width=0.98\textwidth,height = 3.8cm]{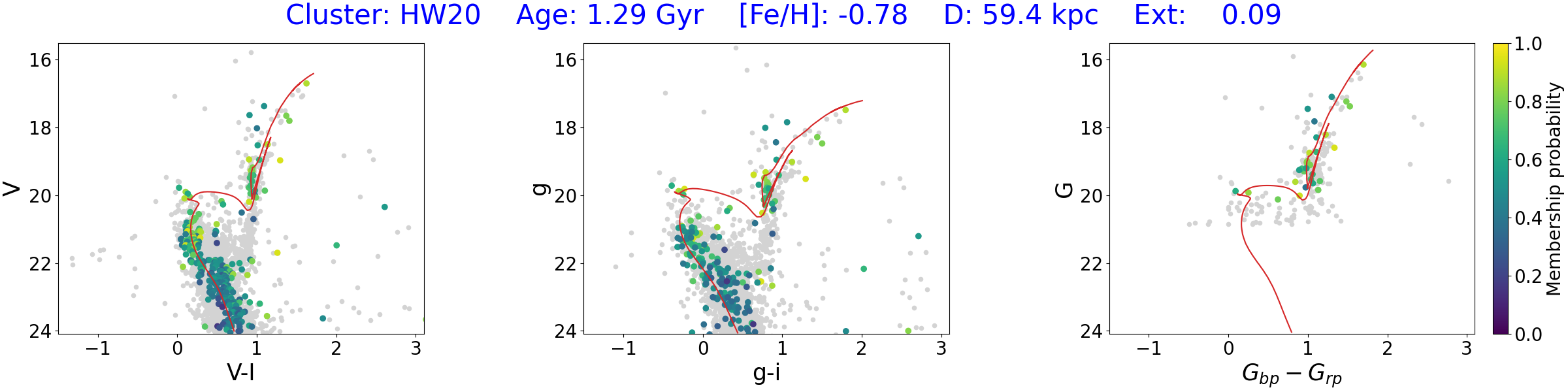}
    \end{subfigure} 
    
    \begin{subfigure}[b]{\textwidth}
         \includegraphics[width=0.98\textwidth,height=3.8cm]{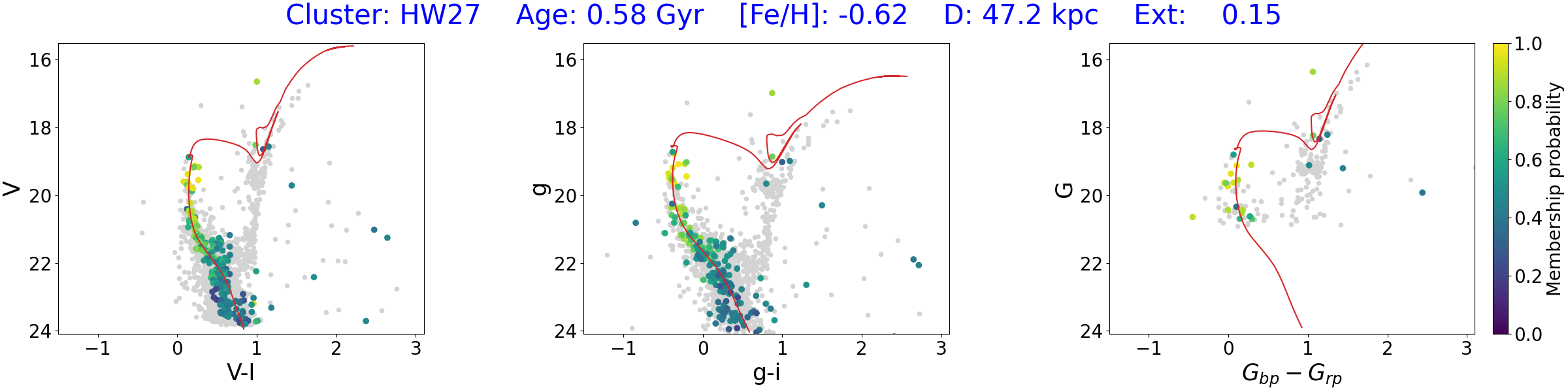}
     \end{subfigure}
    \caption{ Best fitting isochrones for the decontaminated CMDs of all the sample clusters. The CMDs for each cluster from different surveys, including SMASH and Gaia are plotted in their respective filters.
    The membership probability derived for the VISCACHA sample was adopted for the matched stars in other catalogues as well. The red solid line represents the best fitting isochrone. The presence of only the isochrone in a plot signifies that no data were available from the corresponding survey.  }
    \label{MBfit01}
\end{figure*}
\begin{figure*}
    \centering
     \begin{subfigure}[b]{\textwidth}
         \includegraphics[width=0.98\textwidth,height=3.4cm]{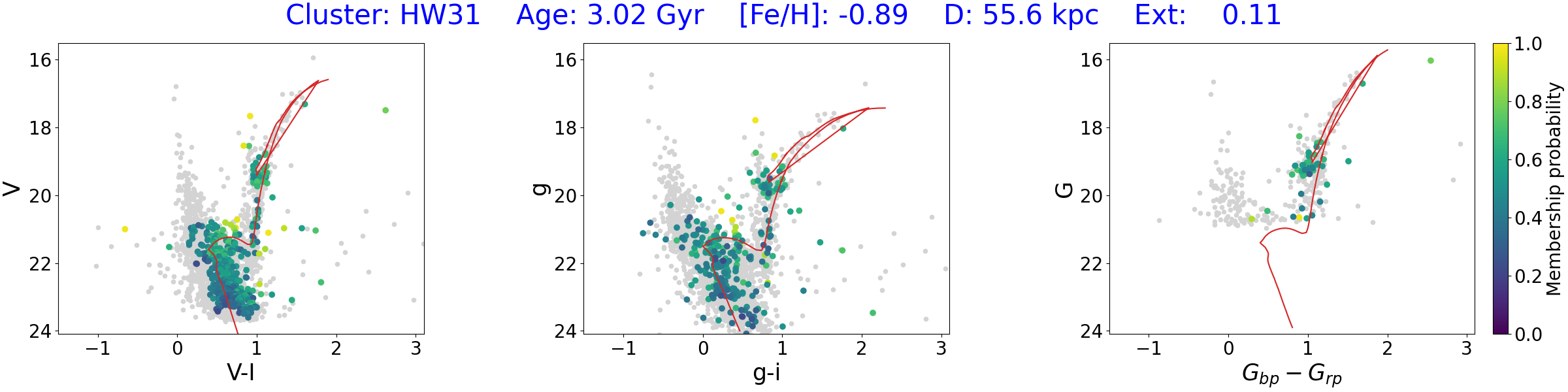}
     \end{subfigure}

     \begin{subfigure}[b]{\textwidth}
         \includegraphics[width=0.98\textwidth,height=3.4cm]{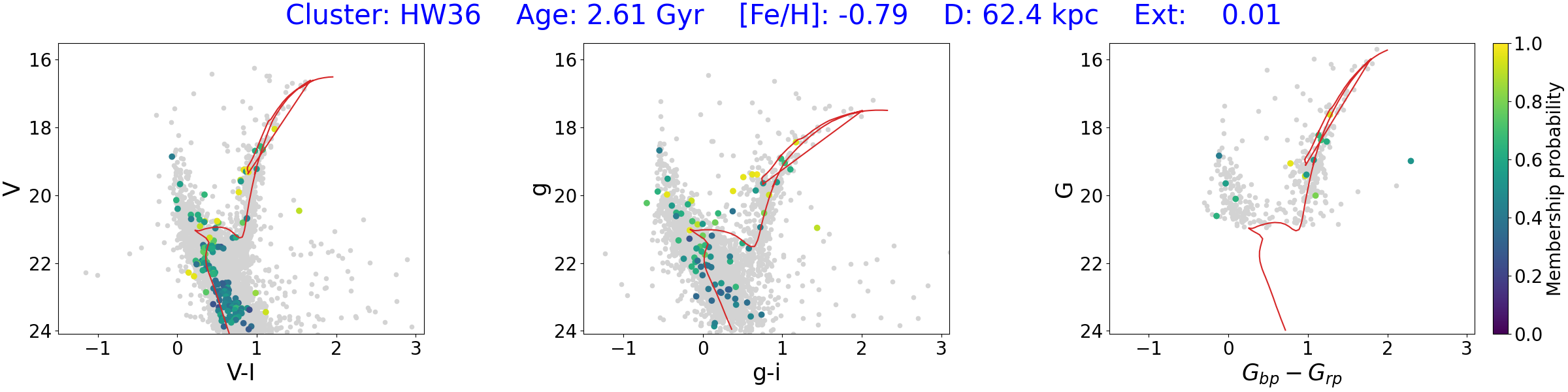}
     \end{subfigure}

      \begin{subfigure}[b]{\textwidth}
         \includegraphics[width=0.98\textwidth,height=3.4cm]{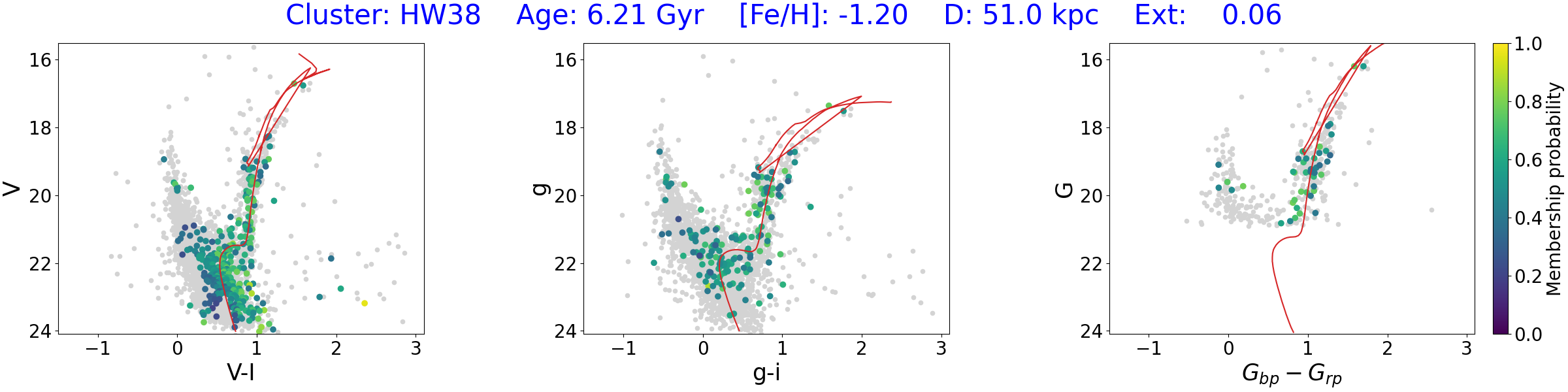}
     \end{subfigure}

    \begin{subfigure}[b]{\textwidth}
        \includegraphics[width=0.98\textwidth,height=3.4cm]{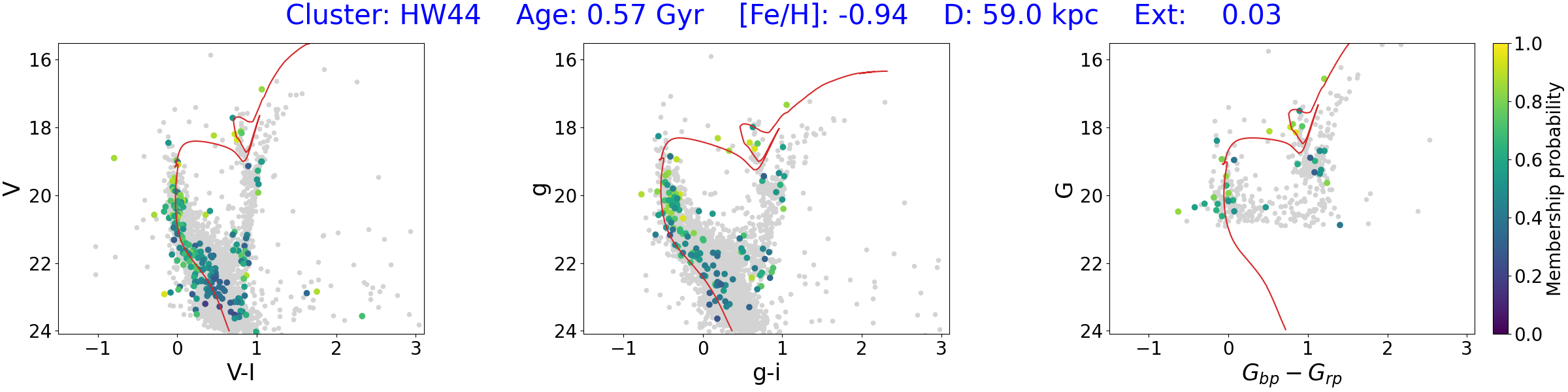}
    \end{subfigure}  

    \begin{subfigure}[b]{\textwidth}
        \includegraphics[width=0.98\textwidth,height=3.4cm]{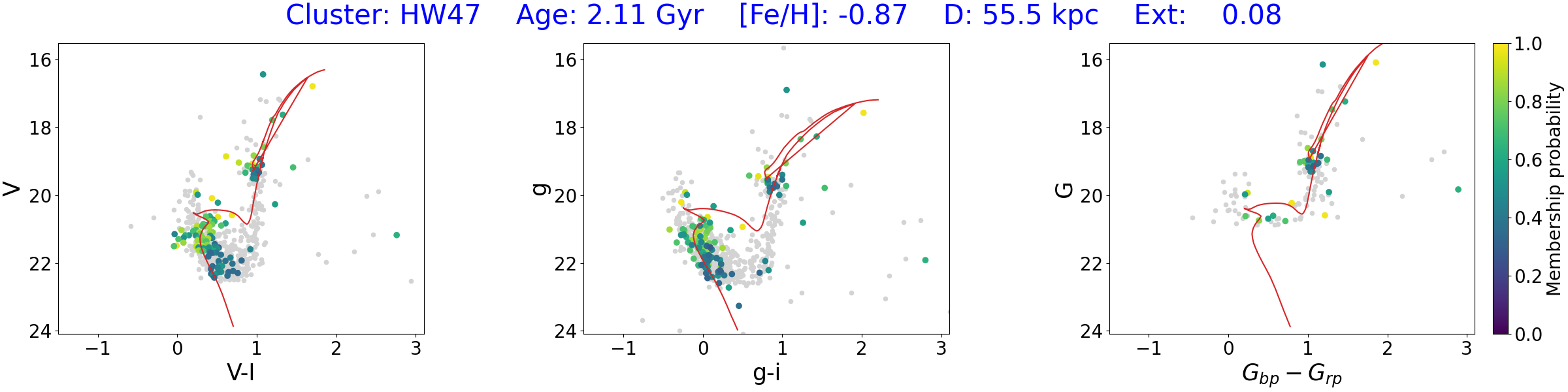}
    \end{subfigure} 
    
    \begin{subfigure}[b]{\textwidth}
         \includegraphics[width=0.98\textwidth,height=3.4cm]{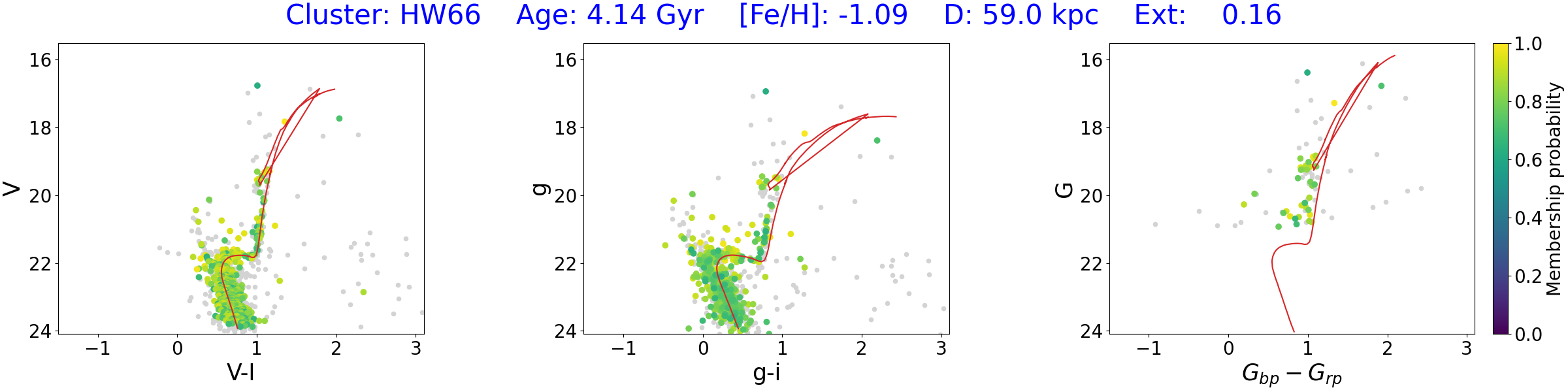}
     \end{subfigure}

      \begin{subfigure}[b]{\textwidth}
         \includegraphics[width=0.98\textwidth,height=3.8cm]{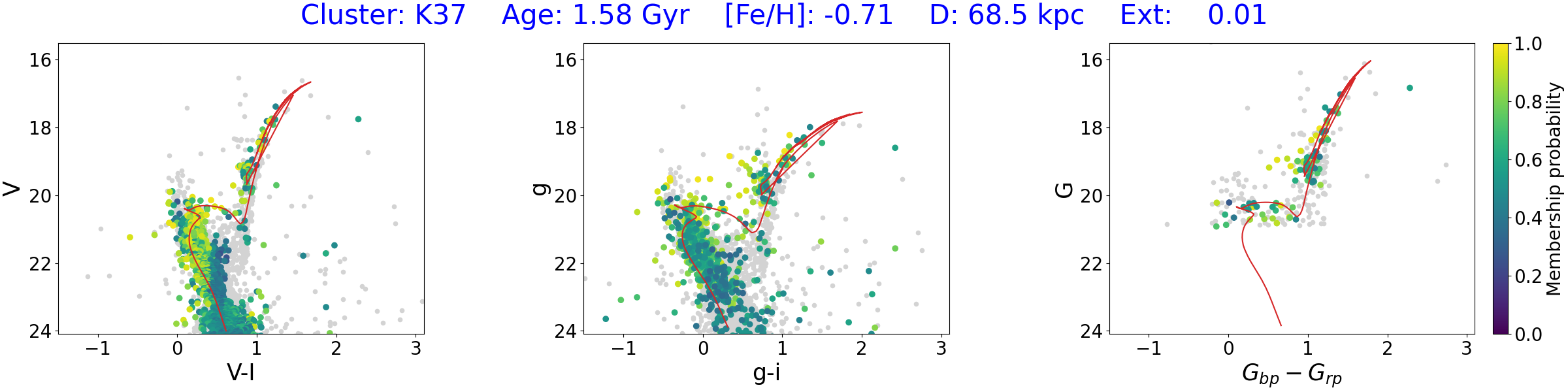}
     \end{subfigure}
    \caption{Same as Fig. \ref{MBfit01} but for more clusters.}
    \label{MBfit02}
\end{figure*}

\begin{figure*}
    \centering
     \begin{subfigure}[b]{\textwidth}
         \includegraphics[width=0.98\textwidth,height=3.4cm]{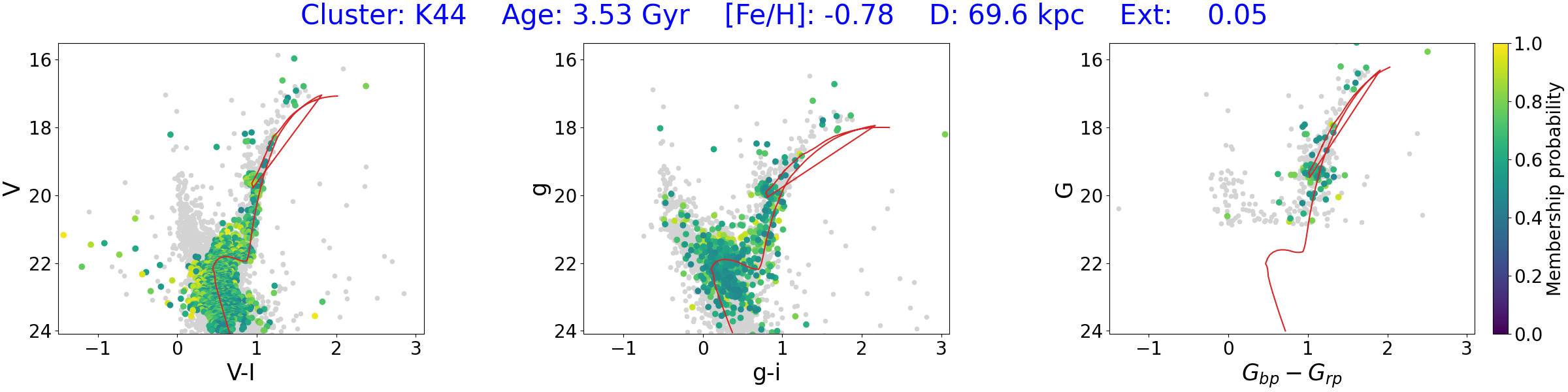}
     \end{subfigure}

      \begin{subfigure}[b]{\textwidth}
         \includegraphics[width=0.98\textwidth,height=3.4cm]{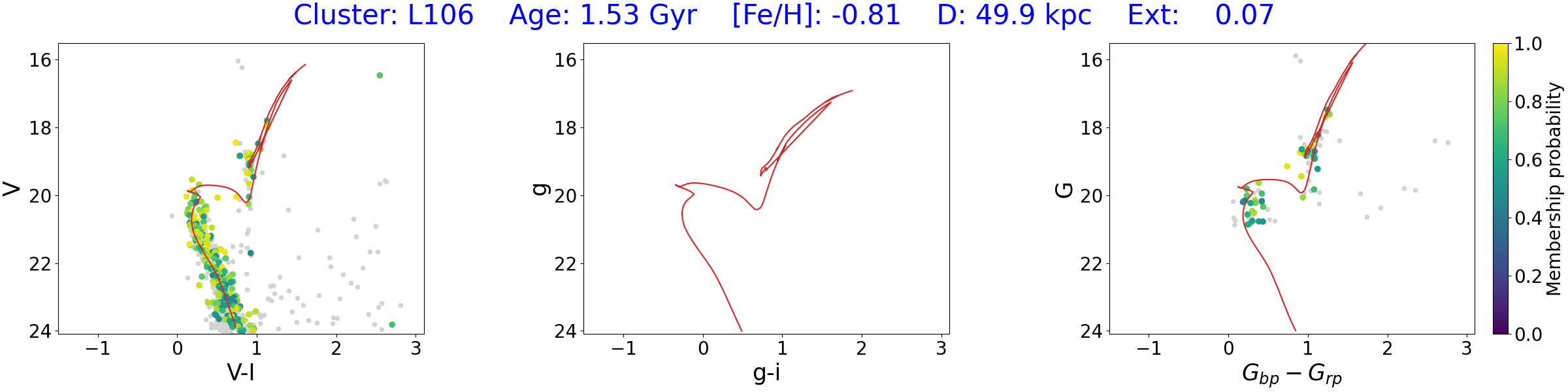}
     \end{subfigure} 
    \begin{subfigure}[b]{\textwidth}
        \includegraphics[width=0.98\textwidth,height=3.4cm]{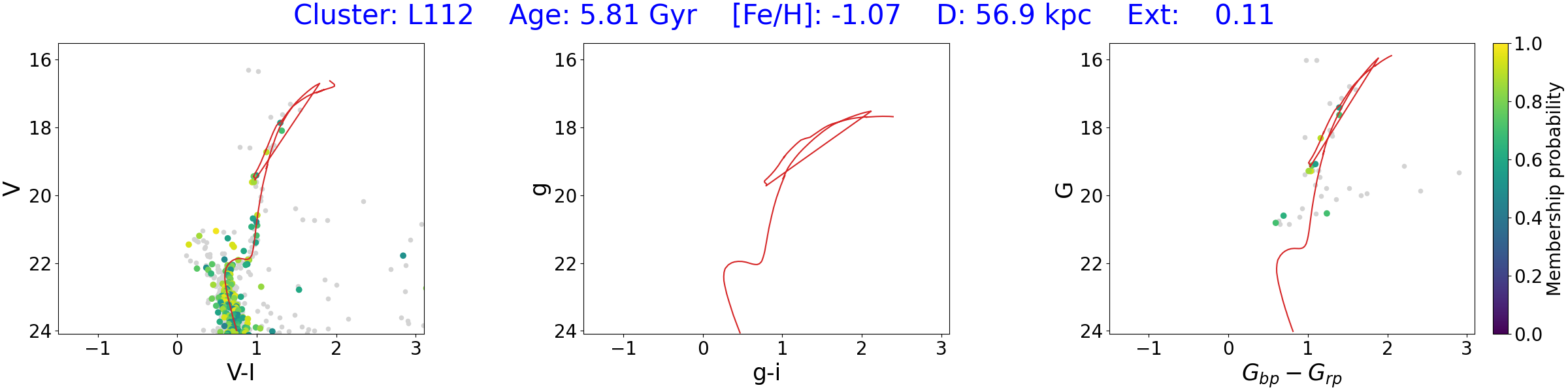}
    \end{subfigure} 
    
    \begin{subfigure}[b]{\textwidth}
         \includegraphics[width=0.98\textwidth,height=3.4cm]{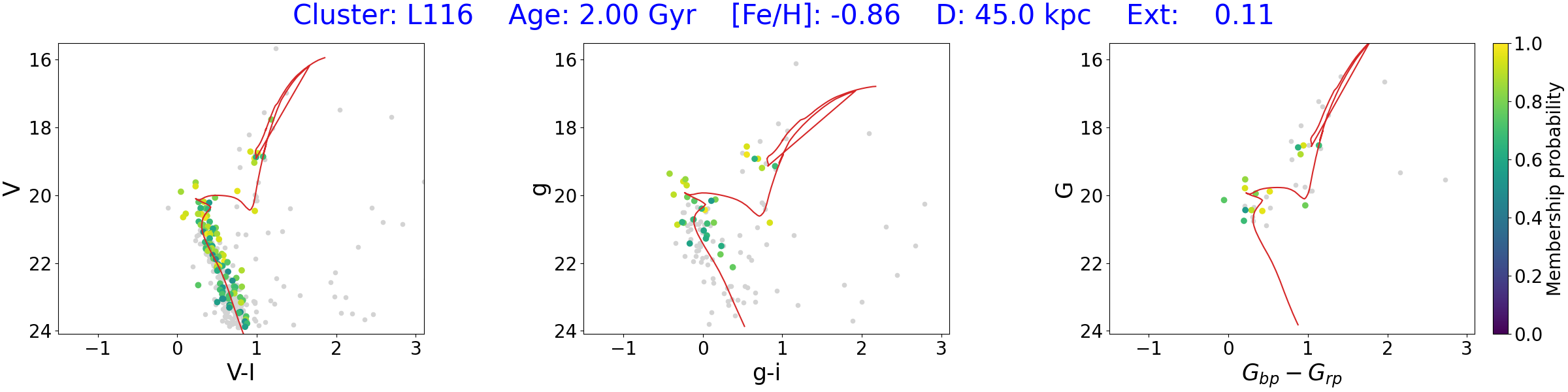}
     \end{subfigure}

      \begin{subfigure}[b]{\textwidth}
         \includegraphics[width=0.98\textwidth,height=3.4cm]{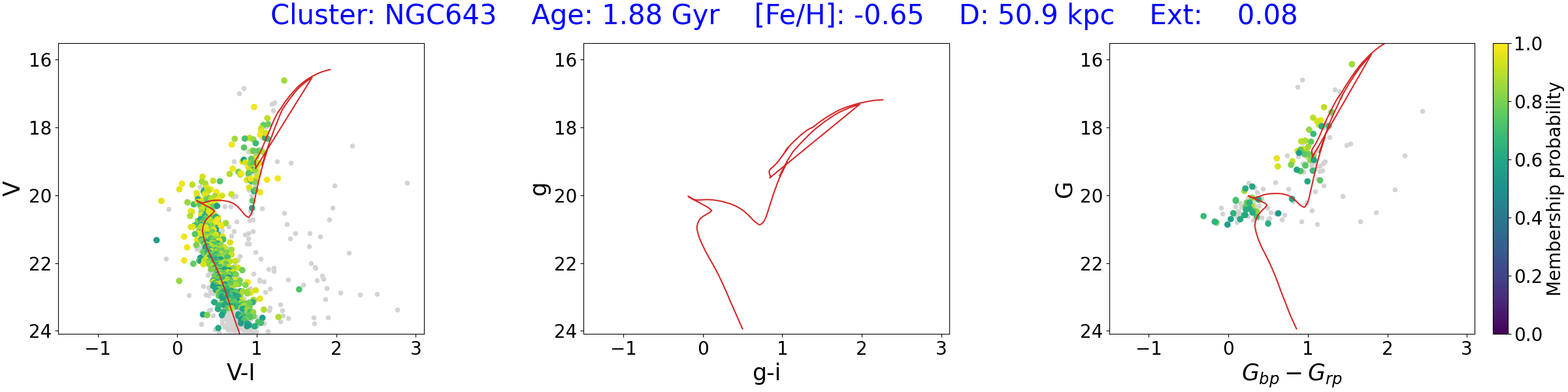}
     \end{subfigure}

     \begin{subfigure}[b]{\textwidth}
         \includegraphics[width=0.98\textwidth,height=3.4cm]{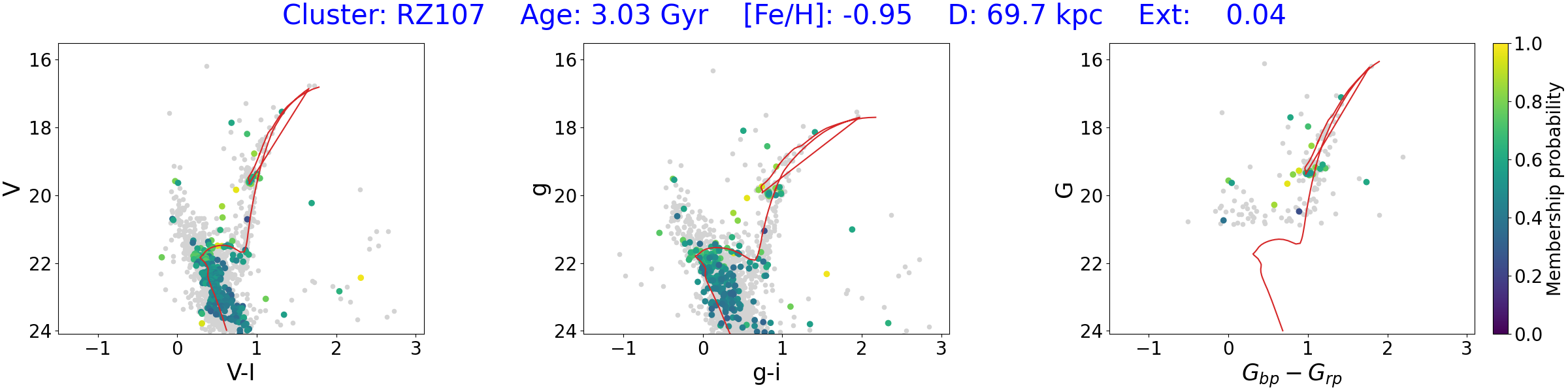}
     \end{subfigure}

      \begin{subfigure}[b]{\textwidth}
         \includegraphics[width=0.98\textwidth,height=3.4cm]{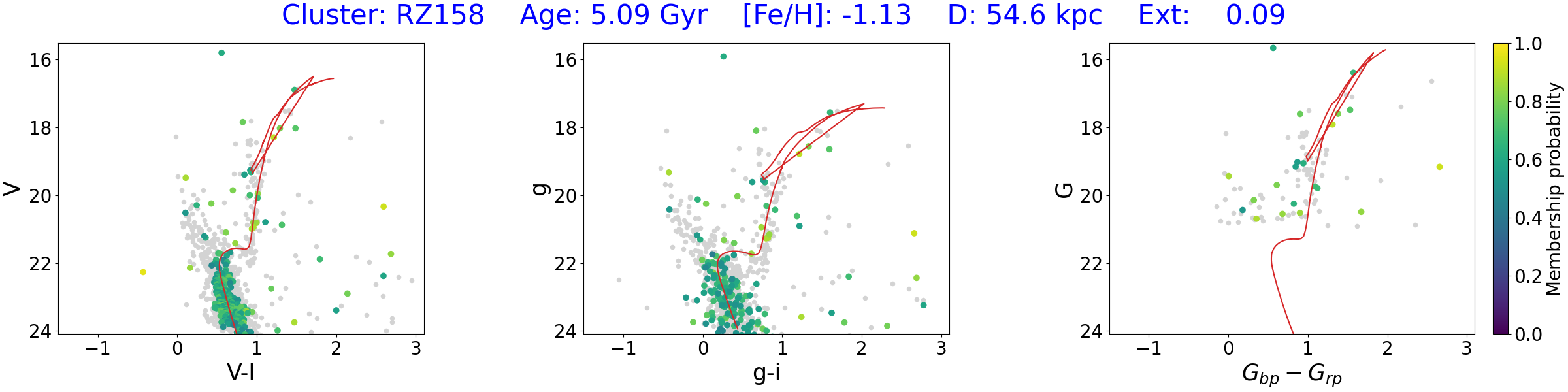}
     \end{subfigure} 
    \caption{Same as Fig. \ref{MBfit01} but for more clusters.}
    \label{MBfit03}
\end{figure*}

\begin{table*}
     \caption{Literature parameters for our cluster sample}
    \label{rescomp}
    \centering
    \footnotesize
    \label{tabcomp}
    \begin{tabular}{lccccc}
    \hline
    \noalign{\smallskip}
    \multicolumn{1}{c}{Cluster} &
    \multicolumn{1}{c}{[Fe/H]} &
    \multicolumn{1}{c}{Age} &
    \multicolumn{1}{c}{E(B-V)} &
    \multicolumn{1}{c}{d} &
    \multicolumn{1}{c}{Reference} \\
    \multicolumn{1}{c}{} &
    \multicolumn{1}{c}{} &
    \multicolumn{1}{c}{(Gyr)} &
    \multicolumn{1}{c}{(mag)} &
    \multicolumn{1}{c}{(kpc)} &
    \multicolumn{1}{c}{} \\
    \hline
    \noalign{\smallskip}
B\,91           & --                      & $0.796\pm0.46$       & $0.24\pm0.10$                     & --   & \citet{Rafelski_Zaritsky2005}\\
                &    --                   & 0.794                         & 0.12                   & --    & \citet{Glatt2010} \\
    \hline 
B\,98ne     & --                      & 1.40$^{+5.77}_{-0.35}$       & --                     & --    & \citet{Rafelski_Zaritsky2005} \\
            &  -0.96$\pm$0.05        & 3.50$\pm$0.50                & 0.09$\pm$0.04         &55.20$\pm$2.50 & \citet{PapIX} \\
\hline
BS\,75          & --                      & 1.26                         &  0.05                  & --    & \citet{Glatt2010}  \\
                & --                      & 1.78                         &  0.00                  & --    & \citet{Maia+2014} \\
                & -0.48$\pm$0.26          & 2.51                         &  0.03$\pm$0.01         & 59.70 & \citet{Perren2017} \\
                &  -0.99$\pm$0.10         & 2.40$\pm$0.20                & 0.10$\pm$0.04         &60.5$\pm$2.50 & \citet{PapIX} \\
\hline
BS\,80          &    --                   & 1.00                         & 0.02                   & --    & \citet{Glatt2010} \\
                &  -0.88$\pm$0.65         & 3.98                         & 0.020$\pm$0.009        & 64.27 & \citet{Perren2017} \\
                &  -1.00$\pm$0.08        & 3.00$\pm$0.50                & 0.08$\pm$0.04         &58.9$\pm$3.30 & \citet{PapIX} \\
\hline
HW\,20          & --                      & 5.69$^{+4.32}_{-0.41}$       & --                     & --    & \citet{Rafelski_Zaritsky2005}\\
                & -0.55$^{+0.13}_{-0.10}$ & 1.10$^{+0.08}_{-0.14}$       & 0.07$^{+0.02}_{-0.01}$ & 62.20$^{+2.50}_{-1.20}$ & \citet{PapI}\\   
    \hline    
HW\,27          & --                      & 0.104       & --                     & --    & \citet{Rafelski_Zaritsky2005}\\
                &    --                   & 0.199                         & 0.05                   & --    & \citet{Glatt2010} \\
    \hline   
HW\,31          & --                      & 1.26                         & 0.57$\pm$0.4          & --    & \citet{Rafelski_Zaritsky2005} \\
                & --                      & 4.60$\pm$0.20                & --                     & --    & \citet{Parisi2014} \\
                &  -0.40$\pm$0.22         & 3.16$\pm$0.05                & 0.05$\pm$0.10         &61.36$\pm$0.80 & \citet{Perren2017} \\
                & -0.89$\pm$0.04          & --                           & --                    & --    & \citet{deBortili2022} \\
                & -0.90$\pm$0.05          & 4.57$\pm$0.63                & --                     & 54.95$\pm$2.50    & \citet{Piatti2025} \\
    \hline
HW\,36          &                         & 1.38$^{+4.90}_{-0.03}$       & --                     & --    & \citet{Rafelski_Zaritsky2005} \\
                & --                      & 1.32$^{+0.72}_{-0.30}$     & 0.09                   & --    & \citet{Gatto2021} \\
                &  -0.84$\pm$0.14         & 2.80$\pm$0.70                & 0.05$\pm$0.06         &58.1$\pm$4.30 & \citet{PapIX} \\
\hline
HW\,38          & --                      & $\sim0.60$       & 0.67$\pm$0.18        & --   & \citet{Rafelski_Zaritsky2005,Glatt2010} \\
                & -0.85                   & 3.89$^{+0.27}_{-1.90}$               & 0.06                   & --    & \citet{Gatto2021} \\
\hline
HW\,44          &  -0.6                     & $0.42\pm0.35$                         & 0.11       & -- & \citet{Gatto2021} \\
\hline
HW\,47          & -0.92$\pm$0.04          & 3.30$\pm$0.50                & --                     & --    & \citet{Parisi2009,Parisi2014} \\
                &  -0.40$\pm$0.22         & 1.58$\pm$0.10                & 0.02$\pm$0.007         &61.9$\pm$0.62 & \citet{Perren2017} \\
                & -0.90$\pm$0.02          & 2.51$\pm$0.15                & --                     & 54.95$\pm$1.50    & \citet{Piatti2025} \\
    \hline
HW\,66          & -1.30/-1.35$\pm$0.20    & 4.00$\pm$0.90/3.50$\pm$1.00  & 0.05                   & --    & \citet{2011Piatti} \\
                & --                      & 3.40$\pm$0.40                & --                     & --    & \citet{Parisi2014} \\
                &  -0.88$\pm$0.65         & 2.82                         & 0.02$\pm$0.01          & 64.27 & \citet{Perren2017} \\
                &  -1.16$\pm$0.04        & 4.00$\pm$0.50                & 0.16$\pm$0.04         &57.50$\pm$2.40 & \citet{PapIX} \\
    \hline       
K\,37           & --                      & 1.00                         & --                     & --    & \citet{Rafelski_Zaritsky2005}\\
                & $-0.79\pm0.11$          & $1.80\pm0.20$                & --                     & --    & \citet{Parisi2015,Parisi2014}\\
                & $-0.58 \pm 0.33$ & $1.12\pm0.07$                       & $0.04\pm0.01$          & $62.52\pm1.3$  & \citet{Perren2017}\\
                & -0.81$^{+0.13}_{-0.14}$ & 1.81$^{+0.24}_{-0.21}$       & 0.05$^{+0.01}_{-0.02}$ & 62.40$^{+2.30}_{-1.80}$ & \citet{PapI}\\
                & -0.80$\pm$0.01          & 1.77$\pm$0.08                & --                     & 53.46$\pm$0.97    & \citet{Piatti2025} \\
\hline
K\,44           &                         & 5.64                         & 0.78$\pm$0.04          & --    & \citet{Rafelski_Zaritsky2005} \\
                &  -0.85$\pm$0.04         & --                & --          & -- & \citet{Parisi2015} \\
                & -0.85                   & 3.71$\pm$0.07                & 0.09                   & --    & \citet{Gatto2021} \\
                &  -0.78$\pm$0.03         & --                & --          & -- & \citet{parisi+22} \\
\hline
L\,106          & -0.88$\pm$0.06          & 2.00$\pm$0.30                & --                     & --    & \citet{Parisi2009,Parisi2014} \\
                &  -0.70$\pm$0.43         & 1.58$\pm$0.10                & 0.00$\pm$0.01          & 59.2$\pm$0.06 & \citet{Perren2017} \\
    \hline     
L\,112          & -1.08$\pm$0.07          & 5.10$\pm$0.30                & --                     & --    & \citet{Parisi2014,Parisi2015} \\
                &  -0.88$\pm$0.65         & 3.98$\pm$0.10                & 0.01$\pm$0.01          &63.09$\pm$1.40 & \citet{Perren2017} \\   
    \hline
L\,116          & -1.10$\pm$0.02          & 2.80$\pm$1.00               & 0.06                   & -- & \citet{Piatti2001} \\
                & -0.89$\pm$0.02          & --                          & --                     & --    & \citet{parisi+22} \\
                & -1.14$\pm$0.08          & 2.45$\pm$0.28                & --                     & 42.27$\pm$3.27    & \citet{Piatti2025} \\
    \hline 
RZ\,107         & --                      & 1.37$^{+2.08}_{-0.01}$       &  --                    & --    & \citet{Rafelski_Zaritsky2005} \\
                &  -0.83$\pm$0.10         & 2.90$\pm$0.20                & 0.06$\pm$0.04         &67.9$\pm$3.40 & \citet{PapIX} \\
\hline
RZ\,158         & --                      & 7.27$^{+2.73}_{-0.18}$       & --                     & --   & \citet{Rafelski_Zaritsky2005} \\
                & -0.90$^{+0.43}_{-0.39}$ & 4.80$^{+1.60}_{-1.30}$       & 0.06                   & 54.70$\pm$3.5 &  \citet{Bica2022}\\
                &  -1.11$\pm$0.05         & 4.70$\pm$1.40                & 0.08$\pm$0.05         &56.00$\pm$4.40 & \citet{PapIX} \\
\hline
NGC\,643         & --                      & 1.50$\pm$0.30               & --                     & -- & \citet{Bica+1986} \\
                & -0.82$\pm$0.03          & 2.00$\pm$0.30               & --                     & --    & \citet{Parisi2009,Parisi2014} \\
    \hline 
    \multicolumn{6}{c}{SB clusters already studied, but not in our sample}\\       
    \hline
HW\,79          & -1.30/-1.40$\pm$0.2     & 5.00$\pm$1.30/4.30$\pm$1.20  & 0.06/0.07              & --    & \citet{2011Piatti} \\
                & --                      & 4.10$\pm$0.50                & --                     & --    & \citet{Parisi2014} \\
                & -0.88$\pm$0.65          & 6.31                         & 0.01$\pm$0.01          & 63.68 & \citet{Perren2017} \\
                & -1.26$\pm$0.03            & 4.9$\pm$0.20                 & 0.08$\pm$0.02            & 56.5$\pm$1.1   & \citet{PapIX} \\
\hline
NGC\,339        & -1.19$\pm$0.10          & --                           & --                     & --    & \citet{Dacosta1998} \\
                & --                      & 6.5$\pm$0.50                 & 0.06/0.04              & 56.2  & \citet{Glatt2008b} \\
                & -1.10$\pm$0.12          & 5.6$\pm$0.08                 & 0.04                   & --    & \citet{Narloch2021} \\
                & -1.24$\pm$0.01          & --                           & --                     & --    & \citet{Mucciarelli2023} \\        
\hline
    \noalign{\smallskip}
    \end{tabular}

\end{table*}

\section{Revisiting the West Halo Clusters with \texttt{SIESTA}} \label{WHCs}
The astrophysical parameters of the 15 WH clusters were derived in \cite{PapVIII} (PapVIII) by visually fitting the isochrones to the VISCACHA CMDs. 
Here we re-derived the astrophysical parameters for these 15 WH clusters, using the \texttt{SIESTA} code. The results show excellent agreement with the visually fitted isochrones, as we use the same data set to derive the parameters. The corner plots generated with SIESTA for the WH clusters, are provided as supplementary material Fig. B1 (see, section~\ref{cornerplots}). This approach not only improves the results in terms of homogeneity, but also statistically verifies our conclusions about the chemical evolution of WH clusters from \citetalias{PapVIII} (see appendix \ref{WHCs} for more information).
Here we present the results of statistical isochrone fitting using the \texttt{SIESTA} code and compare it with clusters from \citetalias{PapVIII}. 

In the correlation plot for $[\text{Fe/H}]$ (left panel in Fig. \ref{corrplot}), the WH cluster NGC121 is the prominent outlier comparing with our estimate from visually fitted isochrones. The best fitted isochrone for NGC121 yields $\text{age} = 9.77\pm0.39$~Gyr, $[\text{Fe/H}] = -1.32\pm0.05$~dex, $\text{D} = 71.1\pm1.4$~kpc, $\text{E(B-V)} = 0.02\pm0.01, \text{BF} = 0.69\pm0.17$ (\citealp{Bernardo+2024}). The large differences in the values for NGC 121 with the \citetalias{PapVIII} results are primarily due to the high crowding in its central region, where non-resolved systems lead to greater dispersion in the CMD. This is particularly evident in the Red Giant Branch and Red Clump, where a large number of unresolved stars cause photometric inaccuracies. The aforementioned values are derived excluding the central 43 arcsec region which reduces this crowding effect. But on the other hand the result is closer to the value derived from high-resolution spectroscopy, $[\text{Fe/H}] = -1.28\pm0.06$~dex (\citealp{Dalessandro16}). The isochrone fits on the HST data yielded an age of $9.70\pm0.75$~Gyr, and $[\text{Fe/H}] = -1.20\pm0.10$~dex, fairly closer to our estimates with \texttt{SIESTA}. 

In the correlation plot for age (right panel in Fig. \ref{corrplot}) we have a few outliers beyond the $1~\sigma$ region, but many lie within the $2~\sigma$ region with a maximum difference of 0.85~Gyr. The only cluster outside the $2~\sigma$ region is HW1 having an age of 5.63~Gyr derived by \texttt{SIESTA}, where as in \citetalias{PapVIII} we derived a value of 4.6~Gyr. 

Another cluster that is not analyzed in this paper is NGC152, already analyzed using \texttt{SIESTA} by \cite{Bernardo+2024}, and the derived values are $\text{age} = 1.41\pm0.06$~Gyr, $[\text{Fe/H}] = -0.85\pm0.11$~dex, $\text{D} = 59.98\pm2.0$~kpc, $\text{E(B-V)} = 0.11\pm0.02, \text{BF} = 0.39\pm0.14 $, and this fairly agrees with the one derived by \citetalias{PapVIII} and references therein. With high resolution spectroscopy \cite{Song+2021} derived a value of $[\text{Fe/H}] = -0.73\pm0.11$~dex, and this will further align it with our model. 

\begin{figure*} [t]
    \centering
         \includegraphics[width=0.49\textwidth,height = 0.5\textwidth]{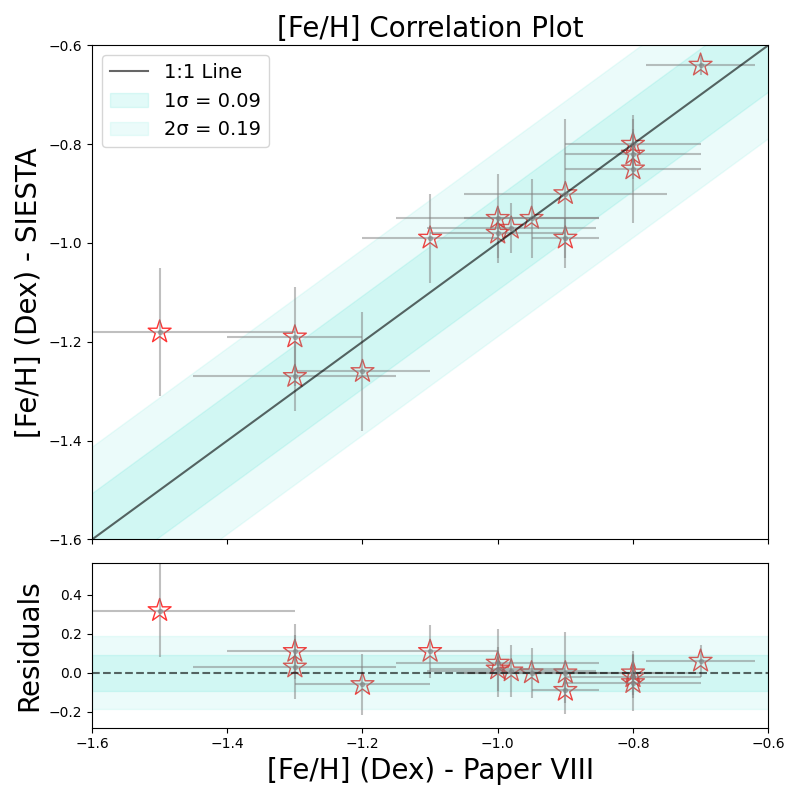}
         \includegraphics[width=0.49\textwidth,height = 0.5\textwidth]{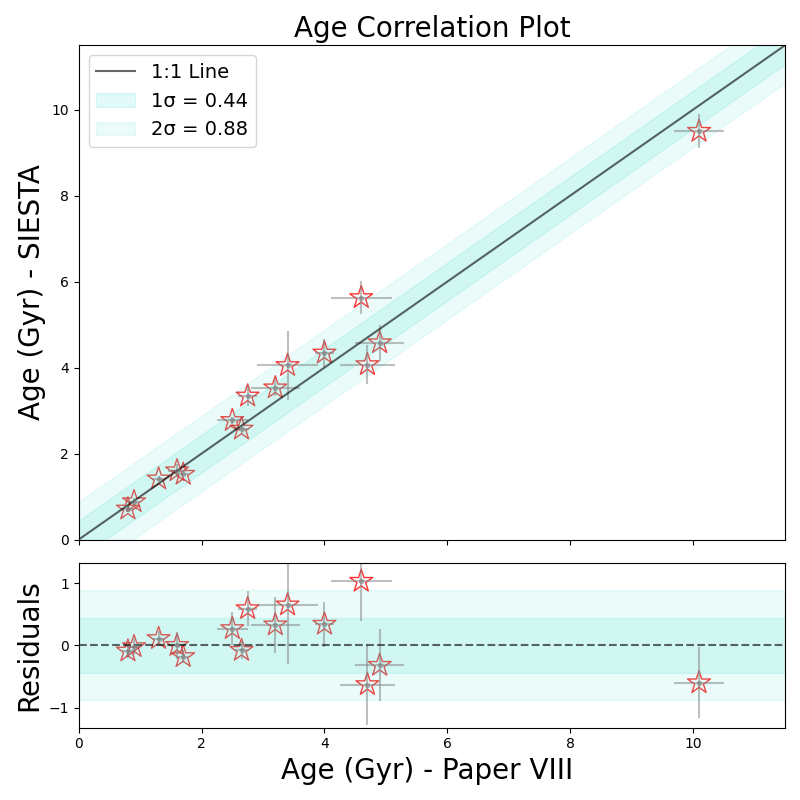}
    \caption{The correlation plots between the metallicities (left) and ages (right) derived using the \texttt{SIESTA} code compared to the values from previous VISCACHA papers, where the ages and metallicities were derived by visually fitting isochrones to the VISCACHA CMDs \citepalias{PapVIII}.
    The corresponding residual plots are displayed beneath each main plot. 1 and 2 sigma regions are shaded to provide a visual representation of the deviations in the correlation plot. }
    \label{corrplot}
\end{figure*}

\begin{table*}
    \centering
    \caption{Astrophysical parameters of the WH clusters re-derived using the \texttt{SIESTA} code.}
    \label{tabresWH}

    \begin{tabular}{p{3cm}p{2.5cm}p{2.5cm}p{2.5cm}p{2.5cm}p{3cm}}
         \hline
         \noalign{\smallskip}
         Cluster Name & Age & [Fe/H] & Distance  & ${\text{E(B-V)}}$  & Binary Fraction  \\
         & [Gyr] &&[kpc]& [mag] & -\\ 
        \noalign{\smallskip}
         \hline\hline
        \noalign{\smallskip}
         
         AM3     &$4.07\pm0.46$      &$-0.99\pm0.09$       & $67.07\pm3.40$       & $0.02\pm0.02$      & $0.12\pm0.07$     \\ \noalign{\smallskip}
         B1     &$0.88\pm0.08$      &$-0.80\pm0.05$       & $65.80\pm2.80$       & $0.02\pm0.01$      & $0.13\pm0.07$     \\ \noalign{\smallskip}
         B2      &$1.54\pm0.02$      &$-0.69\pm0.02$       & $60.57\pm1.65$       & $0.09\pm0.01$      & $0.22\pm0.10$     \\ \noalign{\smallskip}
         B4     &$3.34\pm0.24$      &$-0.97\pm0.05$       & $54.37\pm1.71$       & $0.08\pm0.01$      & $0.13\pm0.10$     \\ \noalign{\smallskip}
         HW1     &$5.63\pm0.39$      &$-1.26\pm0.12$       & $55.76\pm1.68$       & $0.02\pm0.02$      & $0.17\pm0.14$     \\ \noalign{\smallskip}
         HW5     &$4.58\pm0.42$      &$-1.19\pm0.10$       & $58.02\pm1.95$       & $0.09\pm0.02$      & $0.16\pm0.11$     \\ \noalign{\smallskip}
         K6     &$1.52\pm0.03$      &$-0.64\pm0.02$       & $64.07\pm1.50$       & $0.12\pm0.01$      & $0.56\pm0.12$     \\ \noalign{\smallskip}
         K7     &$2.57\pm0.08$      &$-0.95\pm0.08$       & $65.43\pm1.86$       & $0.03\pm0.02$      & $0.22\pm0.14$     \\ \noalign{\smallskip}
         K8     &$2.77\pm0.08$      &$-0.82\pm0.05$       & $63.37\pm2.06$       & $0.01\pm0.01$      & $0.12\pm0.06$     \\ \noalign{\smallskip}
         K9     &$0.71\pm0.06$      &$-0.99\pm0.04$       & $56.44\pm2.45$       & $0.10\pm0.02$      & $0.13\pm0.07$     \\ \noalign{\smallskip}
         K11     &$4.05\pm0.80$      &$-0.95\pm0.09$       & $53.12\pm2.68$       & $0.09\pm0.03$      & $0.21\pm0.16$     \\ \noalign{\smallskip}
         L2     &$4.34\pm0.33$      &$-1.27\pm0.07$       & $56.26\pm1.70$       & $0.07\pm0.02$      & $0.17\pm0.12$     \\ \noalign{\smallskip} 
         L14    &$3.53\pm0.20$      &$-0.98\pm0.05$       & $64.02\pm1.87$       & $0.02\pm0.01$      & $0.60\pm0.22$     \\ \noalign{\smallskip}
         \hline
         \noalign{\smallskip} 
        \multicolumn{6}{c}{WH clusters analyzed in VISCACHA paper X}\\ 
        \noalign{\smallskip} 
        \hline
        \noalign{\smallskip}
         NGC121     &$9.77\pm0.39$      &$-1.32\pm0.05$       & $71.10\pm1.40$       & $0.02\pm0.01$      & $0.69\pm0.17$     \\ \noalign{\smallskip}
         NGC152     &$1.41\pm0.06$      &$-0.85\pm0.11$       & $59.98\pm2.0$       & $0.11\pm0.02$      & $0.39\pm0.14$     \\  
        \noalign{\smallskip}  
         \hline
    \end{tabular}
\end{table*}

\endgroup

\end{document}